\documentclass[12pt]{article}

\usepackage{jheppub}

\usepackage{amsmath,amsfonts}
\usepackage{amssymb}
\usepackage{macros}
\usepackage{caption}
\usepackage{subcaption}
\usepackage{bbm}
\usepackage{axodraw2}
\usepackage{braket}
\usepackage{slashed} 

\renewcommand{\sl}{\slashed}

\newcommand{\ms}{\mathbb} 
\newcommand{\vs}{}	
\newcommand{\gs}{}

\newcommand{\aut}{\operatorname{Aut}}
\newcommand{\R}{\ms R}
\renewcommand{\T}{\ms T}
\renewcommand{\L}{\ms L}
\newcommand{\Q}{\ms Q}
\newcommand{\U}{\ms U}
\renewcommand{\B}{\ms B}
\newcommand{\M}{\ms M}
\renewcommand{\S}{\ms S}
\renewcommand{\Pr}{\ms {P}}
\newcommand{\C}{\ms L}
\renewcommand{\one}{{\mathbbm 1}}

\renewcommand{\xi}{z}

\renewcommand{\p}{{\vs p}}
\renewcommand{\k}{\vs k}

\renewcommand{\c}{c}

\newcommand{\gS}{\gs S}
\newcommand{\gC}{\gs C}
\newcommand{\gF}{\gs F}

\newcommand{\D}{\Delta}

\newcommand{\adj}{\text{adj}}

\def\circrad{4}

\newcommand{\ben}{\begin{enumerate}}
\newcommand{\een}{\end{enumerate}}

\renewcommand{\t}{\tau}

\newcommand{\eps}{\epsilon}

\newcommand{\im}{\operatorname{im}}

\newcommand\tra{^{\mathpalette\raiseT\intercal}}
\newcommand{\ph}{^{\phantom{x}}}

\newcommand\raiseT[2]{%
\setbox0\hbox{$#1{#2}$}\raise\dp0\box0}

\title{Factorization of covariant Feynman graphs for the effective action}

\author[a]{Gero von Gersdorff}
\affiliation[a]{Pontificia Universidade Católica do Rio de Janeiro\\ 
Rua Marquês de São Vicente 225, Rio de Janeiro, Brazil}
\emailAdd{gersdorff@puc-rio.br}

\abstract{We prove a neat factorization property of Feynman graphs in covariant perturbation theory. The contribution of the graph to the effective action is written as a product of a  massless scalar momentum integral that only depends on the basic graph topology, and a background-field dependent piece that contains all the information of spin, gauge representations, masses etc. We give a closed expression for the momentum integral in terms of four graph polynomials whose properties we derive in some detail.
Our results can also be useful for standard (non-covariant) perturbation theory.
}

\begin{document}

\maketitle
\flushbottom

\section{Introduction}

Covariant perturbation theory for the effective action has a long history 
\cite{schwinger1951,dewitt1965dynamical,DeWitt:1967ub,Gilkey:1975iq,Barvinsky:1985an,Avramidi:1990ug,Avramidi:1990je,Fujikawa:1979ay,Fujikawa:1980eg,Ball:1988xg,Schmidt:1993rk,vonGersdorff:2003dt,vonGersdorff:2006nt,Hoover:2005uf,Barvinsky:2005qi,vonGersdorff:2008df}. In recent years there has been some revived interest in the subject \cite{Henning:2014wua,Drozd:2015rsp,delAguila:2016zcb,Henning:2016lyp,Zhang:2016pja,FuentesMartin:2016uol,Ellis:2017jns,Kramer:2019fwz,Ellis:2020ivx,Angelescu:2020yzf,Cohen:2020fcu,Dittmaier:2021fls,Cohen:2023gap,Cohen:2023hmq,Larue:2023uyv}, mostly due to its significance within the Standard Model effective  theory.
With some exceptions \cite{Duff:1975ue,Batalin:1976uv,Batalin:1978gt,Bornsen:2002hh}, covariant perturbation theory has been limited to the one-loop case, where the results can be written as functional traces.

Recently, we have proposed a new covariant formalism based on the heat kernel \cite{Vassilevich:2003xt} which is applicable at any loop order \cite{vonGersdorff:2022kwj}.
In this approach, the effective action is calculated in the background field  (BF) method in a manifestly covariant way.
Recall that in the BF approach, one separates all fields into backgrounds and fluctuations. One then writes down all Feynman graphs to the desired loop order, where the edges of the graphs correspond to BF dependent propagators of fluctuations, and the vertices to BF dependent couplings. Notice that these graphs are connected, one particle irreducible, and have no external lines. 

In this work we will further develop this idea and  show that the contribution to the effective Lagrangian from a given Feynman graph $G$ can  be factorized as follows
\be
\mathcal L^G_{\rm eff}(x)=(-1)^{N_{\rm CFL}}\, {\rm SF} \int d^P\t \
\bigl [ I_{G_0}(\t_i;i\partial_{x_n};\xi_i)
\,\Gamma_G(\t_i;x_n;-i\cev \partial_{\!z_i})
\bigr]_{x_n=x,\ \xi_i=0}\,.
\label{eq:master}
\ee
We will always use an index $i$ to number the edges of the graph ($1\leq i\leq P$), and the index $n$ over its vertices ($1\leq n\leq  V$). 
The $\tau_i$ are  Schwinger parameters, and the $x_n$  spacetime points associated to the vertices of the graph. 
The $\xi_i$  are position variables dual to propagator momenta (in  analogy to the vertex positions $x_n$ being dual to vertex momenta). Working with the $z_i$ variables  greatly simplifies the 
treatment of fermions and derivatives of fields. 
SF denotes the symmetry factor of the graph, and $N_{\rm CFL}$ is the number of closed fermion loops.

\begin{figure}
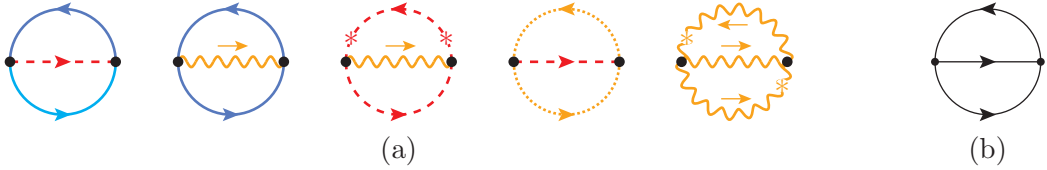

\begin{center}

\begin{subfigure}[b]{0.8\linewidth}
  \centering
  \begin{axopicture}(40,40)(0,0)
    \SetWidth{1.0}
	\LightBlue{\Arc[arrow](20,20)(20,0,180)}
	\Cyan{\Arc[arrow](20,20)(20,180,360)}
	\Red{\DashLine[arrow](0,20)(40,20){3}}
    \SetColor{Black}
	\Vertex(0,20){2}
	\Vertex(40,20){2}

  \end{axopicture}
  \qquad
  \begin{axopicture}(40,40)(0,0)
    \SetWidth{1.0}
    
	\SetColor{LightBlue}
	\Arc[arrow](20,20)(20,0,180)
	\Arc[arrow](20,20)(20,180,360)

	\SetColor{YellowOrange}
	\Photon(0,20)(40,20){2}{6}
	\Line[arrow,arrowpos=0.9,width=0.5,arrowscale=0.8](15,26)(25,26)

    \SetColor{Black}
   	\Vertex(0,20){2}
	\Vertex(40,20){2}


  \end{axopicture}
  \qquad
  \begin{axopicture}(40,40)(0,0)
    \SetWidth{1.0}
    
    \SetColor{Red}
	\DashArc[arrow](20,20)(20,0,180){3}
	\CCirc(2,29){2}{white}{white}
	\Text(2,27){*}
	\DashArc[arrow](20,20)(20,180,360){3}
	\CCirc(38,29){2}{white}{white}
	\Text(38,27){*}
	
	\SetColor{YellowOrange}
	\Photon[arrow](0,20)(40,20){2}{6}
	\Line[arrow,arrowpos=0.9,width=0.5,arrowscale=0.8](15,26)(25,26)
    
    \SetColor{Black}
	\Vertex(0,20){2}
	\Vertex(40,20){2}


\end{axopicture}
\qquad
  \begin{axopicture}(40,40)(0,0)
    \SetWidth{1.0}
	\YellowOrange{\DashArc[arrow](20,20)(20,0,180){1}
	\DashArc[arrow](20,20)(20,180,360){1}}
	\Red{\DashLine[arrow](0,20)(40,20){3}}
    \SetColor{Black}
	\Vertex(0,20){2}
	\Vertex(40,20){2}


  \end{axopicture}
  \qquad
\begin{axopicture}(40,40)(0,0)
    \SetWidth{1.0}

 	\SetColor{YellowOrange}
	\PhotonArc(20,20)(20,0,180){2}{9}

	\CCirc(2,29){2}{white}{white}
	\Text(2,27){*}

	\PhotonArc(20,20)(20,180,360){2}{9}

	\CCirc(38,11){2}{white}{white}
	\Text(38,9){*}
	\Photon[arrow](0,20)(40,20){2}{6}
	
	\Line[arrow,arrowpos=.9,width=0.5,arrowscale=0.8](25,34)(15,34)
	\Line[arrow,arrowpos=.9,width=0.5,arrowscale=0.8](15,26)(25,26)
	\Line[arrow,arrowpos=.9,width=0.5,arrowscale=0.8](15,6)(25,6)

    \SetColor{Black}
	\Vertex(0,20){2}
	\Vertex(40,20){2}


  \end{axopicture}
\caption{}
\end{subfigure}
\begin{subfigure}[b]{0.19\textwidth}
	\centering
	\begin{axopicture}(40,40)(0,0)
    \SetColor{Black}
	\Arc[arrow](20,20)(20,0,180)
	\Arc[arrow](20,20)(20,180,360)
	\Line[arrow](0,20)(40,20)
	\Vertex(0,20){1.5}
	\Vertex(40,20){1.5}
\end{axopicture}
\caption{}
\end{subfigure}
\caption{A number of Feynman graphs (a) that give rise to the common reduced graph (b).
The gauge and Lorentz invariant 2-point functions $\Gamma$ are all different, while the momentum integral $I$ is common to all of them thanks to the same underlying reduced graph. }
\label{fig:dressedreduced}
\end{center}
\end{figure}

The first factor appearing in eq.~(\ref{eq:master}), the function $I_{G_0}(\tau_i;p_n;z_i)$,  is essentially the result of the momentum integral of the graph which we will evaluate in closed form. 
The second one, a gauge-invariant $V$-point function $\Gamma_G(\tau_i;x_n;k_i)$, depends on the background fields and can be readily computed in terms of heat kernel coefficients and field-dependent couplings. 
We will see that the momentum integral $I_{G_0}$ only depends on the topology of the graph $G$. 
Feynman graphs constitute so-called {\em labeled} graphs, which are graphs that carry an identity (or label) at each edge and each vertex. The  labels of the edges are the type of fluctuation fields propagating through them and the vertex labels the corresponding couplings.\footnote{For Feynman graphs the vertex labels are uniquely fixed by the labels of the connecting edges and vice versa.} Notice that several edges (vertices) can carry the same label. While these labels do enter the calculation of $\Gamma_G$,  the calculation of $I_{G_0}$ only depends on the reduced graph $G_0$, which is obtained from the full graph by removing all the labels. 
In particular, all information (mass, spin, and all other quantum numbers such as gauge representations, flavor, extra derivatives etc) can be completely ignored. 
An example is given in figure \ref{fig:dressedreduced}.

This paper is organized as follows. 
We will give in sec.~\ref{sec:Gamma} a brief review of the construction of $\Gamma$, and also describe some of its  properties glossed over in \cite{vonGersdorff:2022kwj}.
Section 3 is devoted to a formal derivation of eq.~(\ref{eq:master}) and some comments about the importance to our findings in standard (non-covariant) perturbation theory.
We devote section \ref{sec:I} to the explicit calculation of $I$ (that is, we will perform the momentum integration in closed form), expressing the result entirely in terms of graph-theoretic quantities in sec.~\ref{sec:char}. This is the most technical section, and and we moved some of the graph theory and linear algebra background into two appendices. In sec.~\ref{sec:sym} we discuss how symmetries of the graph act on the factors $I$ and $\Gamma$, and in sec.~\ref{sec:conclusions} we give our conclusions.

\section{The gauge-invariant function $\Gamma$}

\label{sec:Gamma}

\begin{table}
\begin{eqnarray*}
\raisebox{-20pt}{
  \begin{axopicture}(130,40)(-15,-20)
    \SetWidth{1.0}
    \SetColor{Black}
    \Vertex(0,0){2}
    \Vertex(100,0){2}
    \DashLine[arrow](100,0)(0,0){4}
    %
    \Text(0,15)[t]{$\alpha$}
    \Text(100,15)[t]{$\beta$}
    \Text(-5,0)[r]{$x_n$}
    \Text(105,0)[l]{$x_m$}
  \end{axopicture}
}
 &=&e^{-\tau_im_i^2}\left\{B(\tau_i,X_s;x_n,x_m)\right\}^{\alpha}_{\  \beta}
\\
\raisebox{-20pt}{
  \begin{axopicture}(130,40)(-15,-20)
    \SetWidth{1.0}
    \SetColor{Black}
    \Vertex(0,0){2}
    \Vertex(100,0){2}
    \Line[arrow](100,0)(0,0)
    %
    %
    \Text(50,-10)[t]{$k_i$}
    \Text(0,15)[t]{$\alpha$}
    \Text(100,15)[t]{$\beta$}
    \Text(-5,0)[r]{$x_n$}
    \Text(105,0)[l]{$x_m$}
  \end{axopicture}
}
 &=&e^{-\tau_im_i^2}(i\sl D_n+\sl k_i+m)\left\{B(\tau_i;X_f;x_n,x_m)\right\}^{\alpha}_{\  \beta}
\\
\raisebox{-20pt}{
  \begin{axopicture}(130,40)(-15,-20)
    \SetWidth{1.0}
    \SetColor{Black}
    \Vertex(0,0){2}
    \Vertex(100,0){2}
    \Photon(0,0)(100,0){3}{8}
%
    \Text(0,15)[t]{$\mu,\, a$}
    \Text(100,15)[t]{$\nu,\, b$}
    \Text(-5,0)[r]{$x_n$}
    \Text(105,0)[l]{$x_m$}
  \end{axopicture}
}
 &=&-\left\{B(\tau_i;X_v;x_n,x_m)\right\}^{\mu,\,a}_{\ \ \nu,b}
\end{eqnarray*}
\caption{The basic propagators. See eq.~(\ref{eq:defB}) for the formal definition of $B$.}
\label{tab:prop1}
\end{table}

\begin{table} 
\begin{eqnarray*}
\raisebox{-20pt}{
  \begin{axopicture}(130,40)(-15,-20)
    \SetWidth{1.0}
    \SetColor{Black}
    \Vertex(0,0){2}
    \Vertex(100,0){2}
    \DashLine[arrow](100,0)(0,0){4}
    %
    \CCirc(19,0){\circrad}{white}{white}
    \Text(19,-2.75){\Large{\bf{*}}}
    \Text(19,-10)[t]{$\rho$}
    \Text(50,-10)[t]{$k_i$}
    \Text(0,15)[t]{$\alpha$}
    \Text(100,15)[t]{$\beta$}
    \Text(-5,0)[r]{$x_n$}
    \Text(105,0)[l]{$x_m$}
  \end{axopicture}
}
 &=&e^{-\tau_im_i^2}{(D_{n,\rho}-i k_{i,\rho})}
 	\left\{B(\t_i;X_s;x_n,x_m)\right\}^{\alpha}_{\  \beta}
\\
\raisebox{-20pt}{
  \begin{axopicture}(130,40)(-15,-20)
    \SetWidth{1.0}
    \SetColor{Black}
    \Vertex(0,0){2}
    \Vertex(100,0){2}
    \DashLine[arrow](100,0)(0,0){4}
    %
    \CCirc(81,0){\circrad}{white}{white}
    \Text(81,-2.75){\Large{\bf{*}}}
    \Text(81,-10)[t]{$\rho$}
    \Text(50,-10)[t]{$k_i$}
    \Text(0,15)[t]{$\alpha$}
    \Text(100,15)[t]{$\beta$}
    \Text(-5,0)[r]{$x_n$}
    \Text(105,0)[l]{$x_m$}
  \end{axopicture}
}
 &=&e^{-\tau_im_i^2}{(D_{m,\rho}+ik_{i,\rho})}
 	\left\{B(\tau_i;X_s;x_n,x_m)\right\}^{\alpha}_{\  \beta}
\\
\raisebox{-20pt}{
  \begin{axopicture}(130,40)(-15,-20)
    \SetWidth{1.0}
    \SetColor{Black}
    \Vertex(0,0){2}
    \Vertex(100,0){2}
    \Photon(0,0)(100,0){3}{8}
    \Line[arrow, arrowpos=1](60,10)(40,10)
    \CCirc(19,0){\circrad}{white}{white}
    \Text(19,-2.75){\Large{\bf{*}}}
    \Text(19,-10)[t]{$\rho$}
    \Text(50,-10)[t]{$k_i$}
    \Text(0,15)[t]{$\mu,\, a$}
    \Text(100,15)[t]{$\nu,\, b$}
    \Text(-5,0)[r]{$x_n$}
    \Text(105,0)[l]{$x_m$}
  \end{axopicture}
}
 &=&-{(D_{n,\rho}-ik_{i,\rho})}
 	\left\{B(\tau_i;X_v;x_n,x_m)\right\}^{\mu,\,a}_{\ \ \nu,b}
\\
 \raisebox{-20pt}{
  \begin{axopicture}(130,40)(-15,-20)
    \SetWidth{1.0}
    \SetColor{Black}
    \Vertex(0,0){2}
    \Vertex(100,0){2}
    \Photon(0,0)(100,0){3}{8}
    \Line[arrow, arrowpos=1](60,10)(40,10)
    \CCirc(81,0){\circrad}{white}{white}
    \Text(81,-2.75){\Large{\bf{*}}}
    \Text(81,-10)[t]{$\rho$}
    \Text(50,-10)[t]{$k_i$}
    \Text(0,15)[t]{$\mu,\, a$}
    \Text(100,15)[t]{$\nu,\, b$}
    \Text(-5,0)[r]{$x_n$}
    \Text(105,0)[l]{$x_m$}
  \end{axopicture}
}
 &=&-{(D_{m,\rho}+ik_{i,\rho})}
 	\left\{B(\tau_i;X_v;x_n,x_m)\right\}^{\mu,\,a}_{\ \ \nu,b}
\end{eqnarray*}
\caption{Propagators of derivative of fields. See eq.~(\ref{eq:defB}) for the formal definition of $B$.}
\label{tab:prop2}
\end{table}

We define the $V$-point function $\Gamma_G$ for a given graph $G$  as follows. 
\begin{itemize}
\item
For each edge of $G$ write the expressions given in tables~\ref{tab:prop1} and \ref{tab:prop2}. We refer to these expressions somewhat loosely as {\em propagators}, even though they are only part of the latter.
The basic propagators (without derivatives) are given by the expressions in table \ref{tab:prop1}.
Covariant derivatives of fields are indicated by a star and are subject to the rules
 given in table \ref{tab:prop2}.
Two or more derivatives (on either end of the propagator) are included in an obvious way, and so are derivatives on fermion propagators (the latter do not appear in renormalizable theories).
\item
For each vertex, write the appropriate field-dependent coupling $C_n(x_n)$, obtained from 
differentiation of the interaction $\mathcal L_{\rm int}$ with respect to the fluctuations.\footnote{Notice that we define the contribution of a vertex to $\Gamma$ as $C$ and not $iC$. The remaining factor of $i^V$ will be included in the definition of $I$ instead, see eq.~(\ref{eq:Idef1}).\label{foot:vertexi}}  
We stress that  derivatives only appear in the propagators, never in the couplings.
\end{itemize}

Hence, $\Gamma$ is the product of all of these factors, 
\be
\Gamma_G(\t_i;x_n;k_i)=\prod_{i=1}^P A_i \prod_{n=1}^V C_n\,,
\label{eq:Gamma}
\ee
where $A_i$ are the expressions in tables \ref{tab:prop1} and \ref{tab:prop2}.
Gauge and Lorentz indices  have to be contracted as dictated by the graph; the result is a total gauge and Lorentz singlet. Notice that there is no integration hidden in eq.~(\ref{eq:Gamma}).

A comment on the arrows is in order. Edges corresponding to propagators involving the momentum $k$ need an orientation. For complex fields, we conventionally take this to be the direction of particle number flow. For real fields, this direction is arbitrary but needs to be used consistently in the calculation of $\Gamma$ and $I$. Reversing the direction of  
 edge $i$ in effect corresponds to $\xi_i\to -\xi_i$ in $I$ and $k_i\to-k_i$ in $\Gamma$, yielding the same result in eq.~(\ref{eq:master}). The reversal of the arrow of real edges thus constitutes an isomorphism of graphs (see section \ref{sec:sym}).

The $B$ function appearing in the Feynman rules is formally defined as the ratio of two functions\footnote{We use a different convention for $B$ as opposed to ref.~\cite{vonGersdorff:2022kwj}, the correspondence is 
$B(it;x,y)=B_{\scriptsize\text{\cite{vonGersdorff:2022kwj}}}(t;x,y)$.}
\be
B(it,X;x,y)\equiv \frac{\braket{x|e^{-it(D^2+X)}|y}}{\braket{x|e^{-it\partial^2}|y}}\,.
\label{eq:defB}
\ee
It is a matrix valued function analytic in $t$ near $t=0$ (the non-analyticities cancel between numerator and denominator). Here, $X$ is a BF dependent mass that always features a universal piece
$X_{\rm spin}=-S^{\mu\nu}F_{\mu\nu}^a\mathfrak t^a$ where $S^{\mu\nu}$ are the Lorentz generators in the appropriate representation, but may include other contributions. 
In this work we will not be concerned with the evaluation of $B$ (the heat kernel expansion) which has been extensively studied elsewhere \cite{dewitt1965dynamical,DeWitt:1967ub,Gilkey:1975iq,Avramidi:1990je,Fliegner:1997rk,Vassilevich:2003xt}, and a \textsc{Mathematica} notebook has been included in the arXiv version of ref.~\cite{vonGersdorff:2022kwj} for an automated calculation of the expansion.
In the following we will give some useful identities that were ommitted in ref.~\cite{vonGersdorff:2022kwj}.

From the definition eq.~(\ref{eq:defB}) we have for real $\t$
\be
B(\t,X;x,y)^\dagger=B(\t,X^\dagger;y,x)\,.
\ee
For scalar, fermion and vector fields one has, respectively, 
\bea
X_s^\dagger&=&X_s\,,\\
X_f^\dagger&=&\gamma_0X_f\gamma_0\,,\\
X_v^\dagger&=&g^{-1}X_vg\,,
\eea
and therefore
\bea
 B(\t,X_s;x,y)^\dagger&=& B(\t,X_s;y,x)\label{eq:scalarherm}\,,\\
 B(\t,X_f;x,y)^\dagger&=&\gamma_0  B(\t,X_f;y,x)\gamma_0\,,\\
 B(\t,X_v;x,y)^\dagger&=& g^{-1}B(\t,X_v;y,x)g\,.
\label{eq:Brelations}
\eea
For real fields 
we in addition have that $X$ and $B$ are purely real.

Further useful identities can be derived for the Fermion propagator. Since
\be
\frac{i}{i\sl D-m}=(i\sl D+m)\frac{-i}{\sl D^2-m^2}=
\frac{-i}{\sl D^2-m^2}(i\sl D+m)\,,
\ee
the fermion propagator has an alternative form due to the identity
\be
(i{\sl D}_x+\sl k+m)B(\t;X_f;x,y)
=
B(\t;X_f;x,y)(-i\cev{\sl D}_y+\sl k+m)\,.
\label{eq:profermbis}
\ee
This immediately gives the Hermiticity property
\be
\left\{(i\sl D_x+\sl k+m) B(\t;X_f;x,y)\right\}^\dagger
=\gamma^0\, (i  {\sl D}_y+\sl k+m)  B(\t;X_f;y,x)
\,\gamma^0\,.
\label{eq:fermpropreal}
\ee
eqns.~(\ref{eq:profermbis}) and (\ref{eq:fermpropreal}) are valid only under the $k$ and $\t$ integrations.

\section{Factorization of Feynman graphs}

\label{sec:Dirac}

\begin{figure}
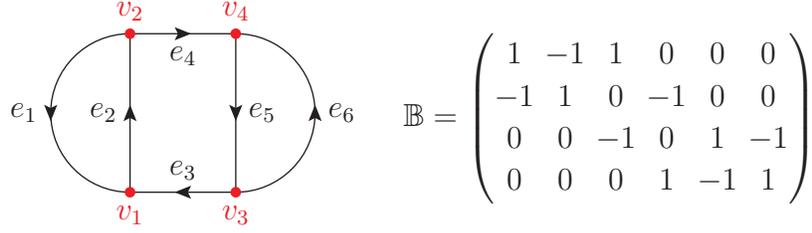

\begin{center}
\be
\raisebox{-35pt}
{
\begin{axopicture}(120,80)(0,0)
\Arc[arrow](40,40)(30,90,270)
\Text(5,40)[r]{$e_1$}
\Line[arrow](40,10)(40,70)
\Text(35,40)[r]{$e_2$}
\Line[arrow](80,10)(40,10)
\Text(60,15)[b]{$e_3$}
\Line[arrow](40,70)(80,70)
\Text(60,65)[t]{$e_4$}
\Line[arrow](80,70)(80,10)
\Text(85,40)[l]{$e_5$}
\Arc[arrow](80,40)(30,270,90)
\Text(115,40)[l]{$e_6$}
\SetColor{Red}
\Vertex(40,10){2}
\Text(40,5)[t]{$v_1$}
\Vertex(40,70){2}
\Text(40,75)[b]{$v_2$}
\Vertex(80,10){2}
\Text(80,5)[t]{$v_3$}
\Vertex(80,70){2}
\Text(80,75)[b]{$v_4$}
\end{axopicture} 
}
\qquad
\B=\begin{pmatrix}
1	&-1	&1	&0	&0	&0\\
-1	&1	&0	&-1	&0	&0\\
0	&0	&-1	&0	&1	&-1\\
0	&0	&0	&1	&-1	&1	
\end{pmatrix}
\nn
\ee
\caption{An example of a three-loop graph and its incidence matrix.}
\label{fig:threeloop}
\end{center}
\end{figure}

In terms of the $V$-point function $\Gamma_G$  defined in the previous section, we get the contribution of the graph $G$ to the effective action   \cite{vonGersdorff:2022kwj}
\be
S^{G}_{\rm eff}=(-1)^{N_{\rm CFL}}\, {\rm SF}
\int d^{P}\!\t 
\int\frac{d^{dP}\!k  }{(2\pi)^{dP}}
\int d^{dV}\!\!x\
\hat I_{G_0}(\t_i;x_n;k_i) \Gamma_G(\tau_i;x_n;k_i)\,,
\label{eq:Seffdirac0}
\ee 
Notice the double integration over both propagator momenta and vertex positions.
We have defined
\be
\hat I_{G_0}(\t_i;x_n;k_i)\equiv i^{V-1-P}\exp\left(k\tra \T_0 k-ix\tra \B_{G_0} k\right)\,.
\label{eq:Idef1}
\ee
The factor of $(-i)^P$ originates from the Wick rotation of the Schwinger parameters $t_i=-i\tau_i$, the factor of $i^V$ from the vertices (see footnote \ref{foot:vertexi}), and the factor of $i^{-1}$ is due to the fact that a Feynman graph $G$ actually computes $i S^G_{\rm eff}$. 
We have furthermore defined
\be
(\T_0)_{ij}\equiv \t_i\delta_{ij}\,,
\ee
as well as the so called incidence matrix of the reduced graph $G_0$, 
\be
(\B_{G_0})_{ni}=\begin{cases}
+1& $if edge $ i $ enters vertex $ n \\
-1& $if edge $ i $ leaves vertex $ n \\
0 & $else$
\end{cases}
\label{eq:Bni}
\ee
which is a $V\times P$ matrix. We adopt the convention that edges which start and end at the same vertex  (self-loops) have a columns of zeros. An example of a graph and its incidence matrix is provided in fig.~\ref{fig:threeloop}.

For the purpose of the following discussion, it is convenient to rewrite eq.~(\ref{eq:Seffdirac0}) in a Dirac notation as
\be
S^G_{\rm eff}=(-1)^{N_{\rm CFL}}\, {\rm SF}
\int d^P\!\t \,
 \braket{\hat I_{G_0}(\tau_i)|\Gamma_G(\tau_i)}\,.
\label{eq:Seffdirac}
\ee 
that is,  
\bea
\hat I_{G_0}(\t_i;x_n;k_n)&=&\braket{\hat I_{G_0}(\t_i)|x_n;k_i}\,,\\
\Gamma_{G}(\t_i;x_n;k_n)&=&\braket{x_n;k_i| \Gamma_{G}(\t_i)}\,.
\eea

In ref~\cite{vonGersdorff:2022kwj} the interior product in eq.~(\ref{eq:Seffdirac}) was computed in the $\ket{x_n;k_i}$ representation. 
One of the main result of the present paper is that it is much more advantageous to compute it in the $\ket{x_n,z_i}$ representation, as the $z_i$ integration turns into a simple derivative, analogous to the $x_n$ integration.
This works because $\Gamma_G(\tau_i;x_n;k_{i})$ is  polynomial in $k_i$. Therefore, Fourier-transforming the latter to the dual $\xi_i$ variables gives a distribution
\be
\braket{x_n;\xi_i|\Gamma_G(\t_i)}=\Gamma_G(\tau_i;x_n;i \partial_{i})\prod_i\delta(\xi_i)\,.
\ee
The momentum-space loop-integration is now entirely contained in 
 $\braket{\hat I_{G_0}|x_n;\xi_i}$ which only depends on the reduced graph $G_0$ and is therefore very universal. We will derive a closed expression for this quantity in sec.~\ref{sec:I}.
To prove eq.~(\ref{eq:master}), consider
$\braket{\hat I_{G_0}(\tau_i)|p_n;\xi_i}$
which will contain  an overall momentum conserving delta function, so we parametrize it in terms of a new function $I_{G_0}$ as
\be
\braket{\hat I_{G_0}(\tau_i)|p_n;\xi_i}=(2\pi)^d\delta(\textstyle\sum_n p_n)I_{G_0}(\tau_i;p_n;\xi_i)\,.
\label{eq:IG0}
\ee
Since we are after the local part of the effective action we expand in the vertex momenta $p_n$. Hence, switching from the $p_n$ to the $x_n$ variables is equally simple:
\be
\braket{\hat I_{G_0}(\tau_i)|x_n;\xi_i}=\int {d}^dx\, 
I_{G_0}(\tau_i;-i\partial_n;\xi_i)\prod_n\delta(x_n-x)\,.
\ee

In summary, both the $x_n$ and $z_i$ integrations have turned into simple derivatives, and in particular eq.~(\ref{eq:Seffdirac}) implies eq.~(\ref{eq:master}).
It remains to compute the function $I_{G_0}$, this is done in the next section; the result can be found in eq.~(\ref{eq:Ifinal}).

The presented method can be modified to work for standard QFT perturbation theory (without   covariant background field method and local derivative expansion). A standard multi-loop Feynman graph gives rise to 
\be
i\mathcal M(p_n)=i^V\!\int \frac{d^{dP}\!k}{(2\pi)^{dP}}(2\pi)^{dV}
\delta
(\p+\B \k)\left[\prod_i \frac{i}{k_i^2-m_i^2}\right] T_G(p_n;k_i)\,,
\ee
where $T_G$ is a tensor (polynomial in the propagator momenta $k_i$), consisting essentially of the nontrivial numerators of the propagators as well as polarization vectors and coupling constants, 
and $p_n$ are  the ingoing  momenta from external lines.\footnote{The momenta of internal vertices (not connected to external lines) are simply zero here.} After introduction of Schwinger parameters, we recognize this as
\be
\mathcal M(p_n)=\int d^P\!\t\!\int \frac{d^{dP}\!k}{(2\pi)^{dP}} \braket{\hat I_{G_0}(\t_i)|p_n;k_i}T_G(p_n;k_i)e^{-\sum_i\t_im_i^2}\,.
\label{eq:standardPT}
\ee
Repeating the same reasoning as above  gives
\be
\mathcal M(p_n)=
(2\pi)^{d}\delta({\textstyle \sum_n p_n})\int d^P\!\t\,
\left[ I_{G_0}(\t_i;p_n;\xi_i)
T_G(p_n;-i\cev\partial_{\xi_i}) e^{-\sum_i\t_im_i^2}\right]_{\xi_i=0}\,,
\label{eq:standardPT2}
\ee
where $I_{G_0}$ is given by eq.~(\ref{eq:Ifinal}).
The formalism thus completely bypasses the usual tensor reduction (e.g.~à la Passarino-Veltman). At this point, let us summarize the main difference to the standard approach:
\begin{itemize}
\item
Using Schwinger instead of Feynman parameters renders the momentum integration in $I_{G_0}(\t_i;p_n;z_n)$ mass-independent. It is noteworthy that the integration over the variable $\sum_i\t_i$ is always trivially possible at the end of the calculation, effectively returning to Feynman parametrization {\em after} the loop momentum integration.
\item
Switching from the $k_n$ to the $\xi_n$ integration in the convolution in eq.~(\ref{eq:standardPT}) results in the explicit factorization eq.~(\ref{eq:standardPT2}) and renders the loop momentum integration independent of $T_G$.
It also sidesteps the parametrization of the propagator momenta $k_i$ in $T_G$ in terms of loop and external momenta. 
\end{itemize}
The remaining integration over the Feynman parameters is  more difficult compared to the effective action (as one does not expand in external momenta). Powerful mathematical methods have been developed over the years to deal with this problem, see \cite{Weinzierl:2022eaz} and references therein.

\section{The universal momentum integral}
\label{sec:I}

In this section we will perform the loop integrations exactly, the result will be precisely the function $I_{G_0}$ of eq.~(\ref{eq:master}).  
As shown in sec.~\ref{sec:Dirac}, this function only depends on the reduced graph $G_0$ (whose information is entirely contained in its incidence matrix). In this section and section \ref{sec:char}, "graph" will always refer to the reduced graph, and the subscript $G_0$ is generally omitted.

\subsection{The fundamental spaces of the incidence matrix}

We can write momentum conservation of the propagator momenta $k_i$ and incoming vertex momenta $p_n$ as
\be
p+\B k=0\,.
\label{eq:momcon}
\ee
The rank of $\B$  for a connected graph is $V-1$ (in general $V-C$ for graphs with $C$ connected components). Each column has one $+1$ and one $-1$,\footnote{The only exception to this are self loops, which have a column of all zeroes.} which implies that  $\sum_n \B_{ni}=0$, which is just overall momentum conservation. Another way of saying this is that $(1,1,\dots 1)\tra$ is in the kernel of $\B\tra$ (a.k.a.~the cokernel of $\B$). If the graph is connected, this is the whole cokernel.
The kernel of $\B$ gives the set of edges whose momenta sum to zero for zero external momenta, that is, precisely the loops. The kernel of $\B$ is  referred to as the {\em cycle space} (loop space would be a more  physics-based terminology). Some properties of this space are reviewed in app.~\ref{app:cycle}. The matrix $\B$ has nullity $L$ (the number of loops) and the matrix $\B\tra $ has nullity $C$ (the number of connected components). The equality of row and column rank gives
\be
P-L=V-C
\label{eq:PLVC}
\ee
For a connected graph ($C=1$) this is simply $P-L=V-1$.

\subsection{Loop integrations}
\label{sec:loop}

It remains to calculate the function $\hat I_{G_0}(\tau_i;p_n;\xi_i)=\braket{\hat I_{G_0}(\t_i)|p_n;\xi_i}$. 
In the remainder of this section, we suppress the subscript $G_0$.
Inserting a complete set of $\ket{x_n;k_i}$ states, using eq.~(\ref{eq:Idef1}) and performing the trivial $x_n$ integrations one gets
\be
\hat I(\tau_i;p_n;\xi_i)= i^{V-1-P}\int\frac{d^{dP}\!k}{(2\pi)^{dP}}
(2\pi)^{dV}
\delta
(\p+\B \k)
\exp\left\{ \k\tra  \T_0 \k 
+i\xi\tra \k
\right\}\,.
\label{eq:Idef}
\ee
which features the usual momentum conserving delta functions for each vertex.

The map $\B$ is not surjective, and, unless
 $L=0$ (tree graphs), neither is it injective, hence  the relation eq.~(\ref{eq:momcon}) cannot be inverted (the $k_i$ cannot be written in terms of the $p_n$). This motivates  the introduction of loop momenta $q_1\dots q_L$ and the restriction to some set of independent external momenta such as $p_1\dots p_{V-1}$ (i.e., one parametrizes the kernel and image of $\B$). 
This is of course the usual procedure of  loop integrations. 
While we certainly could proceed this way, we opt for a different approach which does not 
single out any edges or vertices and hence is more symmetric. Besides being more transparent,
it also simplifies the treatment of graph isomorphisms and automorphisms which typically permute vertices and edges. Keeping symmetries manifest in the resulting expressions can be very helpful in many contexts (see e.g.~sec.~\ref{sec:sym}).

The key technical trick that facilitates this manifestly symmetric treatment of vertices and edges is a Gaussian regularization of the delta functions
\be
\delta
(\p+\B \k)=i^{-V}\left({\eps}{\pi}\right)^{-\frac{dV}{2}}\exp
\{\eps^{-1}(\p+\B \k)\tra(\p+\B \k)\}\,,
\ee
which allows us to directly perform the Gaussian integration  over the (Wick-rotated) $k$ momenta without the need for a parametrization in terms of loop momenta.
Defining the symmetric matrix 
\be
\T\equiv \T_0+\eps^{-1}\B\tra \B\,,
\label{T}
\ee
we can express the result as
\begin{multline}
\hat I(\tau_i;p_n;\xi_i)=i^{-1}(4\pi)^{-\frac{{d(P-V)}}{2}}\eps^{-\frac{dV}{2}}(\det \T)^{-\frac{d}{2}}
\\
\exp\left\{
\eps^{-1}\p\tra (\one- \eps^{-1}\B\T^{-1} \B\tra )\p
-i\eps^{-1}\xi\tra \T^{-1}\B\tra \p+\tfrac{1}{4}\xi\tra \T^{-1}\xi
\right\}\,.
\label{res1}
\end{multline}

We now proceed to carefully take the limit $\eps\to 0$.
We expect the following behavior:\footnote{One way to quickly prove these facts is to write
$\T=\T_0^{\frac12}(\one+\eps^{-1}\T_0^{-\frac12}\B\tra \B\T_0^{-\frac12})\T_0^{\frac12}$ and diagonalize the $P\times P$ symmetric matrix 
$\eps^{-1}\T_0^{-\frac12}\B\tra \B\T_0^{-\frac12}$, which is of the form 
$\diag(\frac{\lambda_1}{\eps} \dots \frac{\lambda_{P-L}}{\eps}0\dots 0)$, with $\lambda_i\neq 0$, from which the small-$\eps$ behaviors of $\det \T$ and $\T^{-1}$ follow immediately.
\label{foot:Tinv}
}
\be
\det \T\sim\mathcal O(\eps^{L-P})\,,\qquad  \T^{-1}\sim
\begin{cases}\mathcal O(1)&L> 0\\ \mathcal O(\eps)&L=0
\end{cases}
\ee
We thus expand $\T^{-1}$ in powers of $\epsilon$:
\be
\T^{-1}=\Q+\eps\,\Q'+\eps^2\Q''+\dots
\label{eq:Tinvexp}
\ee
Comparing coefficients of $\eps^{-1}$ in the equation $\T\T^{-1}=\one$ one gets
$\B\tra \B\Q=0$, and hence, since $\B$ and $\B\tra\B$ have the same kernel,%
\be
\B\Q=0\,.
\label{eq:Bkern}
\ee 
At order $\eps^0$, we have
$
\B\tra \B\Q'+\T_0\Q=\one
$
and hence by use of eq.~(\ref{eq:Bkern}),
$\B\tra \B\Q'\B\tra =\B\tra
\label{eq:bla}
$, which implies that $\B\Q'\B\tra$ acts as the identity on $\im \B$. Since it is also zero on  $\ker \B\tra$, we have
\be
\B\Q'\B\tra =\Pr_{\im \B}\,,
\label{eq:pseudoinvright}
\ee
where $\Pr_{\im \B}$ is the projector onto the image of $\B$.
eqns.~(\ref{eq:Bkern}) and eq.~(\ref{eq:pseudoinvright}) are important identities that we will 
use repeatedly in what follows. 

Using eqns.~(\ref{eq:Bkern}) and  (\ref{eq:pseudoinvright}) in  eq.~(\ref{res1}), we see that the terms quadratic in $p$ give
\bea
\eps^{-1}p\tra (\one- \eps^{-1}\B\T^{-1} \B\tra )p&=&\eps^{-1}p\tra \Pr_{\ker \B\tra }p-p\tra \B\Q''\B\tra p+\mathcal O(\eps)\nn\\
&=&(V\eps)^{-1}\left(\textstyle\sum_n p_n\right)^2-p\tra \B\Q''\B\tra p+\mathcal O(\eps)\,.
\eea
In the second line we have used that $\ker \B\tra$ for a connected graph is one-dimensional and spanned by $(\frac1{\sqrt{V}},\frac{1}{\sqrt V},\dots)$.


We can now safely take the limit $\eps\to 0$ in eq.~(\ref{res1}), producing the expected 
delta function as well as some extra finite terms
\be
\hat I(\tau_i;p_n;\xi_i)=(4\pi)^{-\frac{{dL}}{2}}(2\pi)^d\delta
\left(\textstyle\sum_n p_n\right)
\D^{-\frac{d}{2}}
\exp\left\{
\tfrac{1}{4}\xi\tra \Q\,\xi
-i\xi\tra  \R \,p-p\tra \U p
\right\}\,.
\label{eq:res2}
\ee
Here we defined the shorthands
\bea
\D&\equiv&
V^{-1} \lim_{\eps\to 0}\eps^{P-L}\label{eq:defD}\det\T\,,\\
\R&\equiv& \Q'\B\tra\label{eq:defR}\,,\\
\U&\equiv&\B\Q''\B\tra\,,
\eea
and we recall that $\Q$, $\Q'$ and $\Q''$ were defined in eq.~(\ref{eq:Tinvexp}).
In the second term in the exponential in eq.~(\ref{res1}) we have used eq.~(\ref{eq:Bkern}).
From this, we extract the function 
$I$ from eq.~(\ref{eq:IG0}) by dropping the delta function 
\be
I(\tau_i;p_n;\xi_i)=
(4\pi)^{-\frac{{dL}}{2}}
\D^{-\frac{d}{2}}
\exp\left\{
\tfrac{1}{4}\xi\tra \Q\,\xi
-i\xi\tra  \R \,p
-p\tra \U p
\right\}\,.
\label{eq:Ifinal}
\ee

Finally, we can derive two more interesting relations:
\bea
\tr \T_0 \Q&=&L\label{eq:trQT}\,,\\
\R\tra\T_0\R&=&-\U\,,
\label{eq:UR}
\eea
which easily follow from the $\eps$ expansion of  $\T\T^{-1}=\one$ and the  relations (\ref{eq:Bkern}) and (\ref{eq:pseudoinvright}). Eq.~(\ref{eq:UR}) shows that the matrix $\U$ is actually not independent and that the  vertex momenta $p$ enter in $I$ only in the combination $\R p$.

\subsection{Simple examples at tree-level and one-loop}

Although certainly not the main objective of the formalism, it can be applied to tree level graphs, for instance for integrating out heavy fields diagrammatically (as opposed to simply applying the equations of motion). 
Since $\B$ has trivial kernel, $\B\tra\B$ is actually invertible and one has 
\be
\Delta=1\qquad \Q=0\qquad \R=\B^+
\qquad
\U  = -(\B^+)\tra\T_0\B^+\,,
\ee
where $\B^+=(\B\tra \B)^{-1}\B\tra$ (notice that for tree-level graphs, $\B\tra \B$ has full rank and is invertible).

\begin{figure}
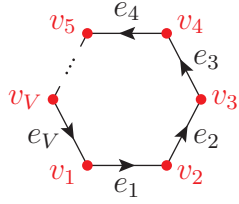

	\centering
	{
	\begin{axopicture}(80,70)(0,0)
		\Line[arrow](25,10)(55,10)
		\Text(40,5)[t]{$e_1$}
		\Line[arrow](55,10)(68,35)
		\Text(65,20)[lc]{$e_2$}
		\Line[arrow](68,35)(55,60)
		\Text(65,50)[lc]{$e_3$}
		\Line[arrow](55,60)(25,60)
		\Text(40,65)[b]{$e_4$}
		\Line(25,60)(22.4,55)
		\Line(14.6,40)(12,35)
		\Vertex(19.8,50){0.5}
		\Vertex(18.5,47.5){0.5}
		\Vertex(17.2,45){0.5}
		\Line[arrow](12,35)(25,10)
		\Text(15,20)[rc]{$e_{V}$}
		\Red{
		\Vertex(25,10){2}
		\Text(25,10)[tr]{$v_1\ $}
		\Vertex(55,10){2}
		\Text(55,10)[tl]{$\ v_2$}
		\Vertex(68,35){2}
		\Text(68,35)[l]{$\ v_3$}
		\Vertex(55,60){2}
		\Text(55,60)[bl]{$\ v_4$}
		\Vertex(25,60){2}
		\Text(25,60)[br]{$v_5\ $}
		\Vertex(12,35){2}
		\Text(12,35)[r]{$v_V\ $}
		}	
	\end{axopicture}}
	\caption{The one-loop polygon graph.}
	\label{fig:polygon}
\end{figure}

At one-loop order, consider the polygon graph from figure \ref{fig:polygon}. One easily finds
\be
\Delta=\sum_i\t_i\qquad \Q_{ij}=\frac{1}{\D}
\label{eq:oneloop1}
\ee
from the definition and the explicit form of the incidence matrix. 
We give the expressions for $\R$ and $\U$ in terms of their polynomials:

\be
\xi\tra \R p=-\sum_{i,j} \frac{\t_i\xi_j}{\D}k_{ij}\,,
\qquad
p\tra \U p=-\sum_{i<j} \frac{\t_i\t_j}{\D}k_{ij}^2\,,
\ee
where
\be
k_{ij}\equiv
\frac{i-j}{V}\sum p_n
+
\operatorname{sgn}(j-i)\!\!\!
\sum_{\min(i,j)+1}^{\max(i,j)}\!\!\! p_n\,,
\ee
(the first term only contributes a total derivative to the effective Lagrangian).

In the next section we will analyze these quantities in the general case.

\section{Characterization of $\Delta$, $\Q$, $\R$ and $\U$}
\label{sec:char}

The quantities $\D$, $\Q$, $\R$ and $\U$ appearing in $I$, eq.~(\ref{eq:Ifinal}),  are entirely encoded in the $\eps$ expansion of $\det \T$ and $\T^{-1}$, which provide a simple and efficient way of calculating them directly, this is likely the preferred method for a possible future automation of the formalism.
In this section we collect some of their properties. The quantities $\D$ and $p^T\U p$ are in fact well known. $\D$ is called the first Szymanzik polynomial, and $\Delta(p^T\U p+ \sum_i m_i^2\tau_i)$ the second Szymanzik polynomial. Some of their graph theoretic properties were studied in \cite{Bogner:2010kv}, see also ref.~\cite{Weinzierl:2022eaz}. The polynomial $\Delta\xi^T\Q\xi$ was studied in \cite{Golz_2017} with the objective of simplifying calculations in quantum electrodynamics. 

\subsection{Graph theory basics}
\label{sec:char1}

\begin{figure}
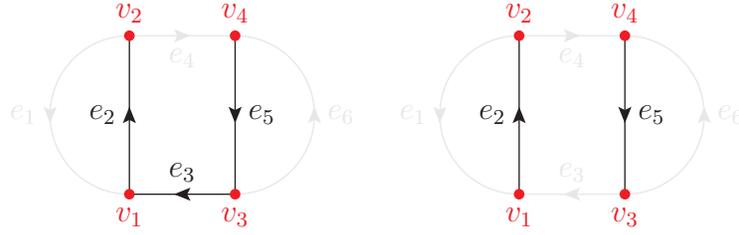

\begin{center}
\be
\raisebox{-35pt}
{
\begin{axopicture}(120,80)(0,0)
\LightGray{\Arc[arrow](40,40)(30,90,270)
\Text(5,40)[r]{$e_1$}}
\Line[arrow](40,10)(40,70)
\Text(35,40)[r]{$e_2$}
\Line[arrow](80,10)(40,10)
\Text(60,15)[b]{$e_3$}
\LightGray{\Line[arrow](40,70)(80,70)
\Text(60,65)[t]{$e_4$}}
\Line[arrow](80,70)(80,10)
\Text(85,40)[l]{$e_5$}
\LightGray{\Arc[arrow](80,40)(30,270,90)
\Text(115,40)[l]{$e_6$}}
\SetColor{Red}
\Vertex(40,10){2}
\Text(40,5)[t]{$v_1$}
\Vertex(40,70){2}
\Text(40,75)[b]{$v_2$}
\Vertex(80,10){2}
\Text(80,5)[t]{$v_3$}
\Vertex(80,70){2}
\Text(80,75)[b]{$v_4$}
\end{axopicture} 
}
\qquad
\raisebox{-35pt}
{
\begin{axopicture}(120,80)(0,0)
\LightGray{\Arc[arrow](40,40)(30,90,270)
\Text(5,40)[r]{$e_1$}}
\Line[arrow](40,10)(40,70)
\Text(35,40)[r]{$e_2$}
\LightGray{\Line[arrow](80,10)(40,10)
\Text(60,15)[b]{$e_3$}}
\LightGray{\Line[arrow](40,70)(80,70)
\Text(60,65)[t]{$e_4$}}
\Line[arrow](80,70)(80,10)
\Text(85,40)[l]{$e_5$}
\LightGray{\Arc[arrow](80,40)(30,270,90)
\Text(115,40)[l]{$e_6$}}
\SetColor{Red}
\Vertex(40,10){2}
\Text(40,5)[t]{$v_1$}
\Vertex(40,70){2}
\Text(40,75)[b]{$v_2$}
\Vertex(80,10){2}
\Text(80,5)[t]{$v_3$}
\Vertex(80,70){2}
\Text(80,75)[b]{$v_4$}
\end{axopicture} 
}
\nn
\ee
\caption{A spanning tree and a spanning 2-forest for the graph of figure \ref{fig:threeloop}.}
\label{fig:threeloopST}
\end{center}
\end{figure}

Let us first define a few more graph theoretic terms.
A {\em spanning tree} of a connected graph ($C=1$) is a connected tree subgraph  that contains all vertices.
It is easily shown that a spanning tree always has $L$ edges less than the original graph. 
Removing further edges from a spanning tree, the graph becomes disconnected ($C'>1$), in this case one obtains so-called {\em spanning $C'$-forests} (spanning 2-forests, spanning 3-forests etc).
Examples of a spanning tree and a spanning forest for the graph of figure \ref{fig:threeloop} are shown in figure \ref{fig:threeloopST}.

\begin{figure}
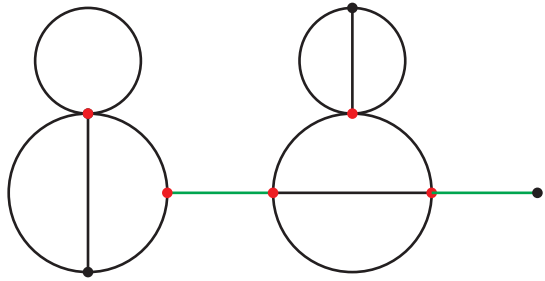


	\centering

 	\begin{axopicture}(200,100)(0,0)
	    \SetWidth{1.0}
		\Arc(30,30)(30,0,360)
		\Green{\Line(60,30)(100,30)}
		\Line(30,0)(30,60)
		\Arc(130,30)(30,0,360)
		\Line(100,30)(160,30)
		\Arc(30,80)(20,0,360)
		\Arc(130,80)(20,0,360)
		\Line(130,60)(130,100)
		\Red{\Vertex(60,30){2}}
		\Red{\Vertex(100,30){2}}	
		\Vertex(30,0){2}	
		\Vertex(30,60){2}	
		\Red{\Vertex(160,30){2}}
		\Red{\Vertex(30,60){2}}
		\Red{\Vertex(130,60){2}}
		\Vertex(130,100){2}
		\Green{\Line(160,30)(200,30)}	
		\Vertex(200,30){2}			
	\end{axopicture}
	\caption{A graph with cut-edges (in green) and cut-vertices (in red). }
	\label{fig:cuts}
\end{figure}

A {\em cut-vertex} is a vertex that, if removed, splits the graph into two disconnected subgraphs with non-empty edge sets.\footnote{The cut vertex belongs to both subgraphs.} 
A {\em cut-edge} or {\em bridge} is an edge that, if removed, splits the graph into two disconnected subgraphs. All edges of a tree graph are bridges.
Notice that the two vertices connected to a bridge are also cut-vertices (unless they are not connected to any other edge). 
A graph containing cut-edges and cut-vertices is shown in figure \ref{fig:cuts}.

Moreover, {\em edge cuts} are sets of edges that, if removed, disconnect the graph.
Bridges are edge cuts with just one element.
A minimal edge cut (i.e. none of its proper subsets is an edge cut) is called a {\em bond}.
For instance, the edges $\{e_1,e_2,e_3\}$ and $\{e_3,e_4\}$ in figure \ref{fig:threeloop} are bonds, and so are any two edges of the polygon graph in figure \ref{fig:polygon}.


Finally, define an equivalence relation on edges as follows: For each edge, let $\mathcal C(e)$ be the set of all cycles $c$ for which $e\in c$. 
Define $e_1\sim e_2$ iff $e_1=e_2$ or $\mathcal C(e_1)=\mathcal C(e_2)\neq \emptyset$.
Loosely speaking, such edges always occur together in cycles.
We show in appendix \ref{app:cycle} that the following definition is equivalent: 
$e_1\sim e_2$ iff $e_1=e_2$ or $\{e_1,e_2\}$ form a bond.  We call two equivalent edges   {\em allies} and the equivalence classes alliances. \footnote{These classes do not appear to be a common topic in graph theory and we have not found any standard terminology for them. In ref.~\cite{Weinzierl:2022eaz} alliances were called chains, this term has however a different meaning in the math literature.}
The edges 3 and 4 in the graph in figure \ref{fig:threeloop} form an alliance, and so do all edges of the polygon graph in figure \ref{fig:polygon}.

\subsection{Explicit expressions 
in terms of subgraphs}
\label{sec:explicit}

Let us define the so-called {\em edge-based Laplacian}  
\be
\C\equiv \B\tra \B.
\ee 
Our approach has to be contrasted with the usual one in terms of the (vertex-based) Laplacian $\B\B\tra$.
Since we will not be making use of the latter in this work we will refer to $\C$ simply as the Laplacian in what follows.

Moreover denote by $\C_{(\ell_1\dots \ell_K)}$, with ($\ell_1<\ell_2<\dots<\ell_k$) 
the Laplacians with rows and columns $\ell_1\dots \ell_K$ deleted. 
These matrices are themselves the  Laplacians of the subgraphs obtained by removing the indexed edges from the original graph.
Directly from the definition eq.~(\ref{T}), we can then write the explicit but slightly cumbersome  expression for $\T^{-1}$:
\be
\T^{-1}=
\frac{\eps(\C+\eps\T_0)^{\rm adj}}{\det(\C+\eps\T_0)}
=\frac{
\sum_{K=0}^{P}
\eps^{K+1} \sum_{\ell_1<\ell_2<...<\ell_K}\t_{\ell_1}\cdots\t_{\ell_{K}}\overline{(\C_{(\ell_1\dots \ell_{K})})^{\rm  {adj}}}
 }
{\sum_{K=0}^P
\eps^{K}\sum_{\ell_1<\ell_2<...<\ell_K}\t_{\ell_1}\cdots\t_{\ell_{K}}\det \C_{(\ell_1\dots \ell_{K})} 
}\,.
\ee
Here, adj stands for adjugate matrix (the cofactor matrix), and the bar means to restore the deleted rows and columns and fill them with zeroes. 
As stated above, the matrices $\C_{(\ell_1\dots\ell_K)}$ are Laplacians of subgraphs, and hence the sums over the $\ell_i$ can be interpreted as sums over subgraphs. Let us characterize the type of subgraphs that contribute at a given $K$.
The nullity of $\C$ (the same as the one of $\B$) is $L$, let $L'$ denote the nullity of  $\C_{(\ell_1\dots \ell_{K})}$.
Note that $L'\leq L$ since we are removing edges and potentially opening loops. Another thing that can happen is that 
the graph becomes disconnected, let us denote by $C'$ the number of connected components.
Using the standard relation eq.~(\ref{eq:PLVC}) as well as $V=V'$ then gives 
\be
L'+K=L+(C'-1)\,.
\ee  
By standard linear algebra results, the rank of the  adjugate matrix of an $N$ dimensional square matrix $\M$ is
\be
\operatorname{rank} \M^\adj=\begin{cases}
N&\operatorname{rank} \M=N\\
1&\operatorname{rank} \M=N-1\\
0&\operatorname{rank} \M<N-1\\
\end{cases}
\label{eq:ranks}
\ee 
The first nonzero term in the numerator occurs when $L'=1$, or $K=L-1$, $C'=1$. In the denominator, the first nonzero term occurs for $L'=0$ or $K=L$, $C'=1$. 
Hence, $\Delta$, $\Q$, $\R$ and $\U$ get contributions from the subgraphs listed in table \ref{tab:subgraphs}.
In particular: 
\bea
\D&=&\sum_\gS \omega_{\gS}\frac{\det\C_{\gS}}{V}\,,
\label{eq:Ddef}
\\
\Q&=&
\sum_{\gC}
\frac{\omega_{\gC}}{\D}\frac{\overline {\C_{\gC}^{\rm adj}}}{V}\,,
\label{eq:Qdef}
\\
\R &=&\sum_\gS \frac{\omega_{\gS}}{\D}
\frac{\overline{\C_{\gS}^{\rm  adj}}\B\tra}{V}\,,
\label{eq:Rdef}
\\
\U
&=&\sum_{\gF}
\frac{\omega_{\gF}}{\D}\frac{\{\B\overline{\C_{\gF}^{\rm  adj}}\B\tra
	-\Pr_{\im \B}\det \C_{\gF}\}}
{V}\,.
\label{eq:Udef}
\eea
where
\be
\omega_X\equiv\prod_{i\notin X}\tau_i
\ee
is the monomial containing the $\t_i$ parameters corresponding to the deleted edges.



\begin{table}
\begin{center}
\begin{tabular}{ccccc}
		&$\Delta$	 	&$\Q$		& $\Q',\ \R$		& $\Q'',\ \U$\\
\hline
edges 	& $P-L$		&$P-L+1$		& $P-L$		&$P-L-1$\\
loops	& 0				&1			&0			&0\\
connected comp.	& 1				&1			&1			&2\\
\hline
		& spanning tree &one-loop connected 		&spanning tree  &spanning 2-forest \\	
				&  $\gS$& $\gC$		& $\gS$ & $\gF=T_1\oplus T_2$\\	

\end{tabular}
\caption{The types of subgraphs contributing to the various $\D$, $\Q$, $\R$ and $\U$. They all have the same number of vertices as the original graph, only fewer edges.}
\label{tab:subgraphs}
\end{center}
\end{table}

\subsubsection{The Kirchhoff polynomial $\D$}
\label{sec:kh}

The determinant
\be
\D
	=\frac{1}{V}\sum_{\substack{\text {spanning}\\ \text{trees } \gS}} \omega_\gS \det\C_{\gS}
\label{eq:kh1}	
\ee
is equal to the so-called Kirchhoff polynomial of the graph\footnote{Sometimes the Kirchhoff polynomial is defined as $\Delta'=\prod_i\t_i\sum_{S}\omega_S^{-1}$. The latter may be given as the determinant of any cofactor of the matrix $\B\T_0\B\tra$.
 In this work we will refer to $\Delta$ as the Kirchhoff polynomial.}
\be
\Delta=\sum_{\substack{\text {spanning}\\ \text{trees } \gS}}\omega_\gS\,.
\label{eq:kh2}
\ee

In order to show the equivalence of the two
we can use the Cauchy-Binet formula  to compute the determinant of $\C_S$
\be
\det \C_\gS=\sum_{n=1}^V(\det \B^{(n)}_\gS)^2
\label{eq:CB}
\ee
where $\B^{(n)}_S$ is the square matrix obtained from $\B_S$ by removing the $n$th row.
One can show by induction  that  $\det \B^{(n)}_S=\pm 1$ \cite{bapat2010graphs} and hence by eq.~(\ref{eq:CB}), 
\be
\det \C_S=V\,.
\label{eq:detLap}
\ee
 As a side remark, the proof  implies that it does not matter whether one uses the directed or undirected graph's incidence matrix.

The Kirchhoff polynomial has the following properties
\begin{enumerate}
\item It is a homogeneous polynomial of degree $L$ in the $\tau_i$ parameters.
\item 
All the coefficients of the monomials appearing in $\Delta$ are equal to one.
\item
It is at most linear in the parameters $\t_i$.
\item
The $\t$ parameter of a cut-edge does not appear in $\Delta$. 
\item
\label{prop:cutvert}
If the graph contains a cut vertex, the corresponding $\Delta$ factorizes, $\Delta=\Delta_1\Delta_2$, where $\Delta_1$ and $\Delta_2$ are the respective polynomials of the separated graphs.
\footnote{This includes the special case of a line starting and ending at the same vertex (i.e., a self-loop).}

\item
The $\tau_i$ parameters of an alliance only enter as their sum.\label{prop:chain} 
\item
It is an invariant polynomial with respect to the automorphism group of the reduced graph.
\end{enumerate}

Properties 1-4 are elementary consequences of eq.~(\ref{eq:kh2}). 
To see property 5, notice that the spanning trees $S$ are in one-to-one correspondence with the spanning trees $(S_1,S_2)$ of the subgraphs separated by the cut-vertex.
Property 6 will be shown in sec.~\ref{sec:Q}, and property 7 is clear from 
 eq.~(\ref{eq:kh1}), as the determinant is  invariant for any value of $\eps$, and hence the coefficients of the $\epsilon$ expansion have to be separately invariant.

\subsubsection{The matrix $\Q$}
\label{sec:Q}

The matrix $\Q$ given in eq.~(\ref{eq:Qdef}) contains the Laplacians of subgraphs with 
$P'=P-(L-1)=V$ edges, and hence the corresponding incidence matrices 
$\B_{C}$ are square matrices. Therefore, 
\be
\C_{C}^{\rm adj}=[\B_{C}\tra \B\ph_{C}]^{\rm adj}
=\B_{C}^{\rm adj}(\B_{C}^{\rm adj})\tra\,.
\ee
According to eq.~(\ref{eq:ranks}) the rank of $\B^\adj_C$ must be one.
It is shown in app.~\ref{app:cycle} that
\be
\B_C^{\rm adj}=
v_C
(1,1,\dots 1)\,,
\ee
where the vector $v_{C}$ spans the one-dimensional kernel of $\B_{C}$, that is, it describes the single loop of $C$.
Its components are $\pm 1$ for edges belonging to the cycle and zero otherwise. 
Thus, 
\be
\C_{C}^{\rm adj}
=V v_{C}\ph v_C\tra\,.
\label{loopB}
\ee
Only the relative sign for different $i,j$ matters,  they have opposite signs if they are oppositely oriented ("anti parallel"). Therefore
\be
v_C^iv_C^j
= \begin{cases}
+1& i,j\in C,\ i,j \text{\ parallel} \\
-1& i,j\in C,\ i,j \text{\ anti-parallel} \\
0& i\notin C \text{\ or\ } j\notin C
\end{cases}
\ee
and
\be
\Q=\sum_{\substack{C \text{\ connected,}\\ {\rm one\ loop}}}\!\!
\frac{\omega_C} {\D}  \ \overline{v^{}_{C}v\tra_{C}}\,,
\label{eq:Qvv}
\ee
where, as previously, the bar instructs us to fill missing rows/columns (edges not belonging to the graph  $C$) with zeros.

We notice furthermore that $ \overline{v}_C$ is in the kernel of $\B$, which confirms our previous result eq.~(\ref{eq:Bkern}).

The objects $v_C$ are elements of the so-called cycle space (see App.~\ref{app:cycle}), in particular they are simple cycles (one-loop).
Since there are usually several one loop graphs $C$ that give rise to the same simple cycle $\c$, one can simplify the expression for $\Q$ further by summing over the simple cycles $\c$ instead of the graphs $C$.
One gets
\be
\Q=\sum_{\substack{\c \text{\ simple}\\ \text{cycle}}} \frac{\D_{\c}}{\D}\, {\c}\,\c\tra
\label{eq:Qbb}
\ee
where $\D_\c$ is the Kirchhoff polynomial for the graph obtained from the original one by contracting that cycle to a point. The equivalence of eqns.~(\ref{eq:Qvv}) and (\ref{eq:Qbb}) is proven in appendix \ref{app:cycle}.
The polynomial $\Delta\xi\tra\Q\xi$ has been described previously in \cite{Golz_2017} where it was dubbed the cycle polynomial.

If there are any cut vertices, then the matrix $\Q$ is block diagonal, since any simple cycle can only belong to one of the two subgraphs separated by the cut-vertex.

The representation of $\Q$ in eq.~(\ref{eq:Qbb}) allows for a simple proof of the alliance property of  $\D$ (property \ref{prop:chain} in the list in sec.~\ref{sec:kh}).
Since $\tr\Q\T_0=L$, 
\be
L\Delta = \sum_{\substack{\c \text{\ simple}\\ \text{cycle}}} \D_{\c}\, \c\tra \T_0 \c\,.
\label{eq:LD}
\ee
 For two allied edges there are two possibilities. Either they are both in $\c$ or both absent from it. If they are both in $\c$ they cannot appear in $\D_c$ because by construction the contracted graph does not contain them. In this case they clearly appear as a sum in $\D$ (from the term $\c^T\T_0\c$).
If they do not appear in $\c$, then  consider the contracted graph. The contracted graph again features an alliance with the same two edges, so one can proceed recursively, {\em q.e.d.}
It follows from  the alliance property of $\Delta$ and $\Delta_\c$ that $\Q$ also has the same property.

Let us summarize the essential characteristics of $\Q$:
\begin{enumerate}
\item
$\Q$ is symmetric.
 \item
$\tr \T_0 \Q=L$.
\item
 $\im \Q = \ker \B$, in particular $\operatorname{rank}(\Q)=L$.
\item
$\Q$ can be written as a sum over simple cycles, eq.~(\ref{eq:Qbb}).
 \item
The $\tau$ parameters of alliances only enter as  their sum. 
\item
If $e_i$, $e_j$ are allies then $\Q_{ik}=\pm\Q_{jk}$.   
This is clear from eqns.~(\ref{eq:Qvv}) or (\ref{eq:Qbb}).
\item
If there are any cut vertices, $\Q$ becomes block-diagonal.
\end{enumerate}

\subsubsection{The matrix $\R$}

Starting from eq.~(\ref{eq:Rdef}), we compute $\R$ explicitly as follows.
We first notice that 
\be
\overline{\C^\adj_S}\B\tra
=\overline{\C_S^\adj\B_S\tra}\,.
\ee
For a tree graph such as $\gS$, the matrix $\C_\gS$  is in fact regular,
 $\det \C_S=V$, see eq.~(\ref{eq:detLap}), so one gets
\be
V^{-1}\L_\gS^{\adj}\B_\gS\tra
= \L_\gS^{-1}( \B_\gS)\tra\equiv \B_\gS^+\,,
\ee
and therefore,
\be
\R=\sum_{\substack{\text {spanning}\\ \text{trees } S}}\frac{\omega_S }{\Delta}\overline{\B_S^+}\,.
\label{eq:Rspanning}
\ee

Acting with $\B_S^+$  on $\B_S$ shows that it is a left inverse of $\B_S$ and it therefore satisfies the first relation in eq.~(\ref{eq:MP2}) (with $\Pr_{\im \B_S\tra}=\one$). Since it also satisfies the second relation in eq.~(\ref{eq:MP2})  it is equal to the Moore-Penrose pseudo-inverse (see app.~\ref{sec:MP} for details).

\begin{figure}
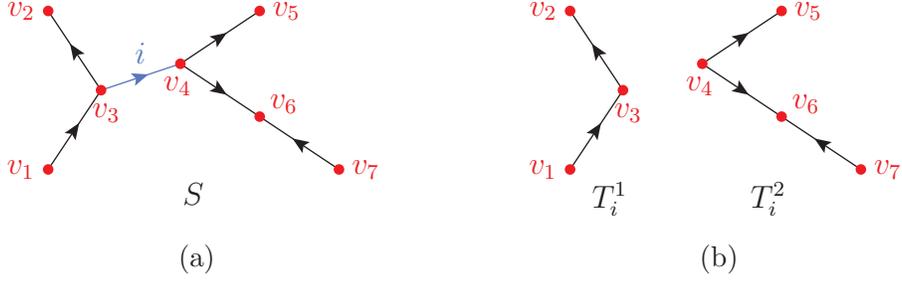

\centering
\begin{subfigure}[b]{0.3\linewidth}
\begin{axopicture}(130,80)(0,-20)
\Line[arrow](10,0)(30,30)
\Line[arrow](30,30)(10,60)
\LightBlue{\Line[arrow](30,30)(60,40)
\Text(45,40)(0)[b]{$i$}}
\Line[arrow](60,40)(90,60)
\Line[arrow](60,40)(90,20)
\Line[arrow](120,0)(90,20)
\SetColor{Red}
\Vertex(10,0){2}
\Text(5,0)(0)[r]{$v_1$}
\Vertex(10,60){2}
\Text(5,60)(0)[r]{$v_2$}
\Vertex(30,30){2}
\Text(32,25)(0)[t]{$v_3$}
\Vertex(60,40){2}
\Text(59,35)(0)[t]{$v_4$}
\Vertex(90,60){2}
\Text(95,60)(0)[l]{$v_5$}
\Vertex(90,20){2}
\Text(94,25)(0)[l]{$v_6$}
\Vertex(120,0){2}
\Text(125,0)(0)[l]{$v_7$}
\Black{
\Text(65,-5)[t]{$S$}}
\end{axopicture}
\caption{}
\label{fig:treea}
\end{subfigure}
\hspace{2cm}
\begin{subfigure}[b]{0.3\linewidth}
\begin{axopicture}(130,80)(0,-20)
\Line[arrow](10,0)(30,30)
\Line[arrow](30,30)(10,60)
\Line[arrow](60,40)(90,60)
\Line[arrow](60,40)(90,20)
\Line[arrow](120,0)(90,20)
\SetColor{Red}
\Vertex(10,0){2}
\Text(5,0)(0)[r]{$v_1$}
\Vertex(10,60){2}
\Text(5,60)(0)[r]{$v_2$}
\Vertex(30,30){2}
\Text(32,25)(0)[t]{$v_3$}
\Vertex(60,40){2}
\Text(59,35)(0)[t]{$v_4$}
\Vertex(90,60){2}
\Text(95,60)(0)[l]{$v_5$}
\Vertex(90,20){2}
\Text(94,25)(0)[l]{$v_6$}
\Vertex(120,0){2}
\Text(125,0)(0)[l]{$v_7$}
\Black{
\Text(25,-5)[t]{$T_i^1$}
\Text(85,-5)[t]{$T_i^2$}
}
\end{axopicture}
\caption{}
\label{fig:treeb}
\end{subfigure}
\caption{The definition of the subgraphs $T_{i}^{1,2}$ appearing in $\R$.}
\end{figure}

It remains to calculate the MP pseudo inverse explicitly.
In view of momentum conservation, $p+\B_S k=0$
\be
 ( \B_S)^+ p=-k\,,
\ee
we expect that $[(\B_S)^+ p]_i$ equals  minus the momentum flowing through propagator $i$.
For a tree level graph such as $S$, this momentum can be found in a rather simple way.
For instance consider the connected tree graph in figure \ref{fig:treea}.
The momentum flowing through edge $i$ can be expressed by either the ingoing or outgoing momenta (recall that all $p_n$ {\em enter} the vertices):
\be
p_{T_i^1}\equiv\sum_{n \in T^1_i} p_n\,,
\qquad
p_{T_i^2}\equiv\sum_{n \in T^2_i} p_n\,,
\label{eq:kS}
\ee
where $T_i^1$ and $T_i^2$ denote the disjoint tree graphs resulting from removing edge $i$ from $S$, where by definition the edge $i$ points from $T_i^1$ to $T_i^2$, see figure \ref{fig:treeb}, i.e. $k_i=p_{T_i^1}=-p_{T_i^2}$.
Any of the two choices  defines a left inverse for $\B_S$, 
and many more are possible by using momentum conservation. We show in app.~\ref{sec:MP} that the one corresponding to the MP inverse is given by
\be
k_{S,i}=-(  \B_S^+p)_i
=
\frac{V_{T^2_i}}{V} p_{T_i^1}-\frac{V_{T^1_i}}{V}p_{T_i^2}
\label{eq:pseuinv}
\ee
Filling in the missing rows of $k_S$ with zeroes using the bar notation, we can write
\be
\R p =-\sum_{\substack{\text {spanning}\\ \text{trees } S}}\frac{\omega_S }{\Delta}
\overline{k_S}\,.
\ee

Let us summarize some of the properties of $\R$:
\begin{enumerate}
\item
$\R$ is a right pseudo-inverse of $\B$, i.e.~$\B\R =\Pr_{\im \B}$, {eq.~(\ref{eq:pseudoinvright}).}
\item 
Notice that $\R$ is not the MP inverse of $\B$, however, $\B^+\equiv\Pr_{\im \B\tra}\R$ is. This follows from the characterization of the MP inverse in  appendix \ref{sec:MP}, see eq.~(\ref{eq:MP3}). 
\item $\R$ satisfies  $\R\Pr_{\ker \B\tra}=0$,
{eq.~(\ref{eq:defR}).}
\item
$\R$ can be written as a sum over spanning trees, eq.~(\ref{eq:Rspanning}).
\item
Replacing $\B_S^+$ by any arbitrary left inverse, for instance $k'_{S,i}=p_{T_i^1}$, just adds a total derivative to the effective Lagrangian. The corresponding function $I'_{G_0}(\t_i;p_n;\xi_i)$ is invariant under graph automorphisms only modulo momentum conservation.
\end{enumerate}

\subsubsection{The matrix $\U$}

The matrix $\U$ is closely related to the so-called second Szymanzik polynomial.
According to table \ref{tab:subgraphs}, it can be written as a sum over spanning two-forests, i.e., tree subgraphs with two connected components containing all the vertices of the original graph. We first observe that the generalization of eq.~(\ref{eq:detLap}) is
\be
\det \mathcal \C_F=\det \C_{T_1}\det\C_{T_2}=V_{T_1}V_{T_2}\,,
\ee
since by a suitable renumbering of the edges we can write $\C_F$ in a block-diagonal form.
Therefore, we can rewrite eq.~(\ref{eq:Udef}) as
\be
\U=
\sum_{\substack{\text {spanning}\\ \text{2-forest }\\ F=T_1\oplus T_2}}
\frac{\omega_{F}}{\D}
\frac{V_{T_1}V_{T_2}}{V}
{\{\overline{\B_F \C_{F}^{-1}\B_F\tra}
	-\Pr_{\im \B}\}}\,.
\ee
We can reexpress it as
\be
\U=
\sum_{\substack{\text {spanning}\\ \text{2-forest }\\ \gF=T_1\oplus T_2}}
\frac{\omega_{\gF}}{\Delta}\frac{V_{T_1} V_{T_2}}{V}\left(
{\overline\Pr}_{\im \B_{T_1}}
+{\overline\Pr}_{\im \B_{T_2}}
-\Pr_{\im \B}
\label{eq:U}
\right)\,.
\ee
In terms of the momenta, this simply becomes
\be
p\tra\U p=-
\sum_{\substack{\text {spanning}\\ \text{2-forest }\\
\gF= T_1\oplus T_2}}
\frac{\omega_{\gF}}{\Delta}
\left(
\frac{ V_{T_2}}{V}p_{T_1}
-\frac{ V_{T_1}}{V}p_{T_2}
\right)^2\,,
\label{eq:pUp}
\ee
where $p_{T_k}=\sum_{n\in T_k} p_n$ is the total momentum influx into $T_k$.
Using momentum conservation, this can also be written as
\be
p\tra\U' p=
\sum_{\substack{\text {spanning}\\ \text{2-forest }\\
\gF= T_1\oplus T_2}}
\frac{\omega_{\gF}}{\Delta}
p_{T_1}\cdot p_{T_2}\,,
\ee
which is the form given in \cite{Bogner:2010kv,Weinzierl:2022eaz}.

We summarize the main properties of $\U$:
\begin{enumerate}
\item
$\U$ is symmetric
\item
 $\Pr_{\ker \B\tra}\,\U=\U\,\Pr_{\ker \B\tra}=0$
\item
$\U=-\R\tra \T_0\R $
\item
$p\tra\U p$ can be given as a sum over spanning 2-forests, eq.~(\ref{eq:pUp}).
\end{enumerate}

\section{Symmetries, symmetry factors, and complex conjugation}
\label{sec:sym}

In this section we would like to investigate the symmetries of a given Feynman graph in relation to our formalism. The purpose is threefold: firstly, we would like to define the symmetry factors of a graph in a more formal way, usually not done in standard quantum field theory textbooks.
Secondly, we would like to study the induced transformations on the $I$ and $\Gamma$ factors and in the case of the latter point out some computational simplification that they imply. Thirdly, we would like to study complex conjugation of graphs in our formalism.

We start by defining {\em isomorphisms} of graphs \cite{Bondy1976}. 
An isomorphism between two  Feynman graphs $G$ and $G'$  is a  label-preserving bijection  $\varphi=(\varphi_v,\varphi_e,\varphi_o)$, where $\varphi_v$ $(\varphi_e)$, maps vertices (edges) of $G$ to vertices (edges) of $G'$, and $\varphi_0$ is a flip of the orientation of some subset of the real edges,  such that 
\be
\S_{\varphi_v}\B_G\,(\S_{\varphi_e})\tra\S_{\varphi_o}=\B_{G'}
\label{eq:Binv0}
\ee
Here, and $\S_{\varphi_{v,e,o}}$ are the  matrix representation of $\varphi_{v,e,o}$. 
The diagonal matrix $\ms S_{\varphi_o}$  has entries $(\ms S_{\varphi_o})_{ii}=-1$ for flipped  edges and $(\S_{\varphi_o})_{ii}=+1$ for all other  edges.
"Label-preserving" simply means that  $\varphi_v$ only maps vertices to vertices of the same kind and similarly for edges. 
Two isomorphic graphs give the same value when applying the Feynman rules.

An {\em automorphism} (or symmetry) of a  graph $G$ is an isomorphism between $G$ and itself. In particular, an automorphism satisfies 
\be
\S_{\varphi_{v}}\B_G\,(\S_{\varphi_e})\tra\ms S_{\varphi_o}=\B_G
\label{eq:Binv}
\ee
Automorphisms form a group, $\aut (G)$,
which is a subgroup  of
 $S_V\times (S_P\ltimes C_2^P)$ where the first factor is the group of vertex permutations, and the second factor the semi-direct product of edge permutations and orientation flips of real edges.
 It is clear that the underlying reduced graph $G_0$ is also invariant under the same operations.  
Notice that if $G$ contains a real self-loop $e_k$, $\aut (G)$ always contains an element that flips the orientation of this edge and does nothing else, according to $\S_{\varphi_v}=\one$, $\S_{\varphi_e}=\one$, and 
$
(\S_{\varphi_o})_{ij} =\delta_{ij}(-1)^{\delta_{ik}}
$.
We define the symmetry factor of the graph as
\be
{\rm SF}=\frac{1}{|\mathcal \aut(G)|}
\ee
that is, the reciprocal of the order of the automorphism group.
Nontrivial automorphism groups are more common for graphs without any external legs.

Notice that $x_n$ and $p_n$ naturally transform under $S_V$,  $\tau_i$ under $S_P$, and $\xi_i$ and $k_i$ transform under $S_P\ltimes C_2^P$. 
Since the graph remains unchanged, we have  that under the action of $\aut(G)$
\bea
\Gamma_G(\tau_i;x_n,k_i)&\to&
\Gamma_{G}(\tau_i;x_n,k_i)
\nn
\\
I_{G_0}(\tau_i;p_n,\xi_i)&\to&
I_{G_0}(\tau_i;p_n,\xi_i)
\eea
In other words, the two factors in our master formula eq.~(\ref{eq:master}) are separately invariant. 
The invariance of $I(\t_i;p_n;\xi_i)$ in eq.~(\ref{eq:Ifinal}) is manifest due to the symmetric treatment of the momenta $k_i$ and $p_n$ in section~\ref{sec:I}.

\begin{figure}
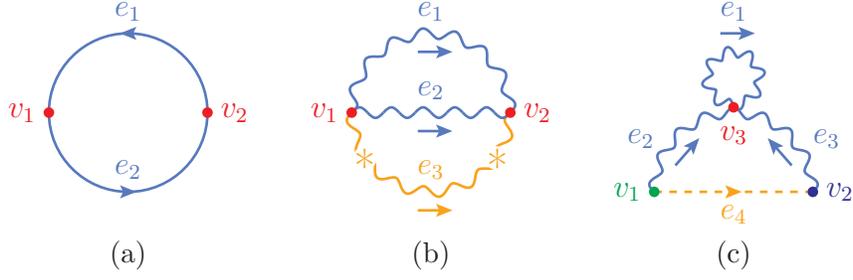

\centering
\begin{subfigure}{0.25\linewidth}
\centering
  \begin{axopicture}(80,80)(0,0)
    \SetWidth{1.0}
    \SetColor{Black}
	\LightBlue{\Arc[arrow](40,40)(30,0,180)}
	\LightBlue{\Arc[arrow](40,40)(30,180,360)}
    \Red{\Vertex(10,40){2}}
    \Red{\Vertex(70,40){2}}
    \Red{\Text(5,40)[r]{$v_1$}}
    \Red{\Text(75,40)[l]{$v_2$}}
    \LightBlue{\Text(40,75)[b]{$e_1$}}
    \LightBlue{\Text(40,15)[b]{$e_2$}}
  \end{axopicture}
  \caption{}
  \label{fig:auto1}
\end{subfigure} 
\begin{subfigure}{0.25\linewidth}
	\centering
	\begin{axopicture}(80,80)(0,0)
   		\SetWidth{1.0}
	    \SetColor{LightBlue}
		\PhotonArc(40,40)(30,0,180){2}{8}
		\YellowOrange{\PhotonArc(40,40)(30,180,360){2}{8}}
		\Photon(10,40)(70,40){2}{5}
		\Line[arrow, arrowpos=1](35,63)(45,63)
		\Line[arrow, arrowpos=1](35,33)(45,33)
		\YellowOrange{\Line[arrow, arrowpos=1](35,3)(45,3)}
    	\Red{\Vertex(10,40){2}}
	    \Red{\Vertex(70,40){2}}
    	\Red{\Text(5,40)[r]{$v_1$}}
	    \Red{\Text(75,40)[l]{$v_2$}}
    	\Text(40,75)[b]{$e_1$}
	    \Text(40,45)[b]{$e_2$}
    	\YellowOrange{\Text(40,15)[b]{$e_3$}
	    \CCirc(15,22.75){\circrad}{white}{white}
    	\Text(15,20.){\Large{*}}
	    \CCirc(65,22.75){\circrad}{white}{white}
    	\Text(65,20.){\Large{*}}}
	\end{axopicture}
	\caption{}
	\label{fig:auto2}
\end{subfigure}
\begin{subfigure}{0.25\linewidth}
	\centering	
	\begin{axopicture}(80,80)(0,0)
		\SetWidth{1.0}
		\LightBlue{
		\PhotonArc(40,10)(30,0,180){2}{8.5}
		\PhotonArc(40,53)(10,0,360){2}{8}
		\Text(40,75)[b]{$e_1$}
		\Line[arrow, arrowpos=1](35,70)(45,70)
		\Line[arrow, arrowpos=1](18,20)(25,28)
		\Line[arrow, arrowpos=1](62,20)(55,28)
		\Text(10,30)[r]{$e_2$}
		\Text(70,30)[l]{$e_3$}
		}
		\Red{
			\Vertex(40,42){2}
			\Text(40,35)[t]{$v_3$}
		}
		\YellowOrange{\DashLine[arrow](10,10)(70,10){3}
		\Text(40,5)[t]{$e_4$}}
		\Green{
			\Vertex(10,10){2}
			\Text(5,10)[r]{$v_1$}
		}	
		\Blue{
			\Vertex(70,10){2}
			\Text(75,10)[l]{$v_2$}	
		}
	\end{axopicture}
	\caption{}
	\label{fig:auto3}
\end{subfigure}
\caption{Some graphs with nontrivial automorphism groups. }
\end{figure}

Let us give some examples. Consider the one-loop graph in Yukawa theory (with a real scalar coupling to fermion-antifermion) in figure \ref{fig:auto1}.
It has a single nontrivial automorphism
given by following transformation (consisting of permuting both the vertices and edges).
\be
\{v_1,v_2;e_1,e_2\}\to   \{v_2,v_1;e_2,e_1\}\,.
\ee
It is easily seen that the graph is unchanged by this operation~\footnote{Notice that it is irrelevant how the graphs are drawn, they are the same and not merely isomorphic.}
\be
\raisebox{-40pt}
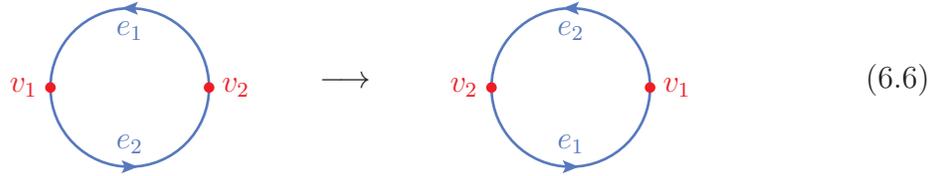
{
  \begin{axopicture}(80,80)(0,0)
    \SetWidth{1.0}
    \SetColor{Black}
	\LightBlue{\Arc[arrow](40,40)(30,0,180)}
	\LightBlue{\Arc[arrow](40,40)(30,180,360)}
    \Red{\Vertex(10,40){2}}
    \Red{\Vertex(70,40){2}}
    \Red{\Text(5,40)[r]{$v_1$}}
    \Red{\Text(75,40)[l]{$v_2$}}
    \LightBlue{\Text(40,65)[t]{$e_1$}}
    \LightBlue{\Text(40,15)[b]{$e_2$}}
  \end{axopicture}
}
\hspace{1cm}
\longrightarrow
\hspace{1cm}
\raisebox{-40pt}{
  \begin{axopicture}(80,80)(0,0)
    \SetWidth{1.0}
    \SetColor{Black}
	\LightBlue{\Arc[arrow](40,40)(30,0,180)}
	\LightBlue{\Arc[arrow](40,40)(30,180,360)}
    \Red{\Vertex(10,40){2}}
    \Red{\Vertex(70,40){2}}
    \Red{\Text(5,40)[r]{$v_2$}}
    \Red{\Text(75,40)[l]{$v_1$}}
    \LightBlue{\Text(40,65)[t]{$e_2$}}
    \LightBlue{\Text(40,15)[b]{$e_1$}}
  \end{axopicture}  
}
\label{eq:graphauto}
\ee 
Since the two graphs are equal they must give the same $\Gamma$. This can be verified by applying the transformation to $\Gamma$ directly:
\bea
&&
\tr\left[C({x_1})(i{\sl D_{ 1}}+{\sl k_1}+m)
B( {\t_1}; {x_1},{x_2})C({x_2})
(i{\sl D_{ 2}}+{\sl \k_2}+m)B( {\t_2}; {x_2},{x_1})
\right]e^{-m^2(\t_1+\t_2)}
\nn\\
&\to&
\tr\left[C({x_2})(i{\sl D_{ 2}}+{\sl k_2}+m)
B( {\t_2}; {x_2},{x_1})C({x_1})
(i{\sl D_{ 1}}+{\sl k_1}+m)B( {\t_1}; {x_1},{x_2})
\right]e^{-m^2(\t_1+\t_2)}\,.
\nn\\
\label{eq:yukreal}
\eea
The two expressions are the same due to the cyclic property of the trace. This is the only 
symmetry and therefore the symmetry factor is $\left|\aut(G)\right|^{-1}=\frac{1}{2}$.

Next, consider the graph from Yang-Mills theory in figure \ref{fig:auto2}.
It has two nontrivial automorphisms, given by
\bea
\{v_1,v_2;e_1,e_2,e_3\}&\to& \{v_2,v_1;-e_1,-e_2,-e_3\}\,,\\
\{v_1,v_2;e_1,e_2,e_3\}&\to& \{v_1,v_2;e_2,e_1,e_3\}\,.
\eea
In the first transformation, the two vertices are permuted and the gauge boson edges are flipped. The graph is indeed invariant under this:
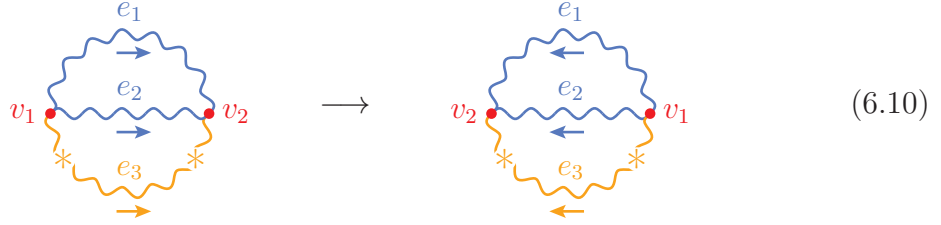
\be
\raisebox{-40pt}{
\begin{axopicture}(80,80)(0,0)
   	\SetWidth{1.0}
    \SetColor{LightBlue}
	\PhotonArc(40,40)(30,0,180){2}{8}
	\YellowOrange{\PhotonArc(40,40)(30,180,360){2}{8}}
	\Photon(10,40)(70,40){2}{5}
	\Line[arrow, arrowpos=1](35,63)(45,63)
	\Line[arrow, arrowpos=1](35,33)(45,33)
	\YellowOrange{\Line[arrow, arrowpos=1](35,3)(45,3)}
    \Red{\Vertex(10,40){2}}
    \Red{\Vertex(70,40){2}}
    \Red{\Text(5,40)[r]{$v_1$}}
    \Red{\Text(75,40)[l]{$v_2$}}
    \Text(40,75)[b]{$e_1$}
    \Text(40,45)[b]{$e_2$}
    \YellowOrange{\Text(40,15)[b]{$e_3$}
    \CCirc(15,22.75){\circrad}{white}{white}
    \Text(15,20.){\Large{*}}
    \CCirc(65,22.75){\circrad}{white}{white}
    \Text(65,20.){\Large{*}}}
	\end{axopicture}
}  
\hspace{1cm}
\longrightarrow
\hspace{1cm}
\raisebox{-40pt}{
\begin{axopicture}(80,80)(0,0)
   	\SetWidth{1.0}
    \SetColor{LightBlue}
	\PhotonArc(40,40)(30,0,180){2}{8}
	\Photon(10,40)(70,40){2}{5}
	\Line[arrow, arrowpos=1](45,63)(35,63)
	\Line[arrow, arrowpos=1](45,33)(35,33)
	
    \Text(40,75)[b]{$e_1$}
    \Text(40,45)[b]{$e_2$}
    
    \YellowOrange{	\PhotonArc(40,40)(30,180,360){2}{8}
	\Line[arrow, arrowpos=1](45,3)(35,3)
    \Text(40,15)[b]{$e_3$}
    \CCirc(15,22.75){\circrad}{white}{white}
    \Text(15,20.){\Large{*}}
    \CCirc(65,22.75){\circrad}{white}{white}
    \Text(65,20.){\Large{*}}
    }
    \Red{\Vertex(10,40){2}}
    \Red{\Vertex(70,40){2}}
    \Red{\Text(5,40)[r]{$v_2$}}
    \Red{\Text(75,40)[l]{$v_1$}}
	\end{axopicture}
}
\label{eq:gbsym}
\ee 
The second transformation consists in a permutation of the first two edges  with the remaining components left unchanged.
The symmetry factor in this case is $\left|\aut(G)\right|^{-1}=\frac{1}{4}$.  

Finally, consider the graph in figure \ref{fig:auto3} from Yang Mills theory with a scalar.
It has a single nontrivial automorphism
\be
\{v_1,v_2,v_3;e_1,e_2,e_3,e_4\}\to
\{v_1,v_2,v_3;-e_1,e_2,e_3,e_4\}\,,
\ee
and hence a symmetry factor of $\frac{1}{2}$.

Computation of symmetry factors is not the only purpose of studying symmetries.
If $\Gamma$ contains terms that are in the same orbit under the symmetry group $\aut (G)$, it suffices to only compute them once.
For instance, in the example of eq.~(\ref{eq:yukreal}), the cross terms 
\begin{multline}
\tr\left[C({x_1})(i{\sl D_{ 1}})
B( {\t_1}; {x_1},{x_2})C({x_2})
{\sl \k_2}B( {\t_2}; {x_2},{x_1})
\right]e^{-m^2(\t_1+\t_2)}\\
\leftrightarrow
\tr\left[C({x_1}){\sl k_1}
B( {\t_1}; {x_1},{x_2})C({x_2})
(i{\sl D_{ 2}})B( {\t_2}; {x_2},{x_1})
\right]e^{-m^2(\t_1+\t_2)}\,,
\end{multline}
transform into each other under the automorphism and only have to be computed once.
This simplification can be quite convenient in sufficiently complicated graphs.

Given a graph $G$ we define its  conjugate graph $G^*$ by reversing all arrows (this may imply the change of vertex labels, see the example below). This transformation acts as complex conjugation on the functions $\Gamma$
and $I$ appearing in eq.~(\ref{eq:master}):
\bea
\Gamma_{G^*}(\t_i;x_n;k_i)&=&\Gamma_G(\t_i;x_n;k_i)^*\label{eq:ccGamma}\,,\\
I_{G_0^*}(\t_i;p_n;\xi_i)&=&I_{G_0}(\t_i;p_n;\xi_i)^*\,.
\eea
The relation for $\Gamma$ follows from the identities given in section \ref{sec:Gamma}, while the relation for $I$ can be seen from eq.~(\ref{eq:Ifinal}).
If the graph $G$ is isomorphic to $G^*$ it is called {\em self-conjugate} or simply {\em real}. 
Real graphs contribute real operators to the effective action. 
Graphs $G$ that are not isomorphic to $G^*$ are called complex, they yield complex operators to the effective action. 

As an example, consider the graph in Yukawa theory with a complex scalar field
\be
\raisebox{-15pt}{
    \begin{axopicture}(80,50)(0,35)
    
    \SetWidth{1.0}
    \SetColor{Black}
	\LightBlue{\DashArc[arrow](40,40)(30,0,180){3}}
	\Cyan{\Line[arrow](10,40)(70,40)}
    \Red{\Vertex(10,40){2}}
    \Green{\Vertex(70,40){2}}
    \Red{\Text(5,40)[r]{$v_1$}}
    \Green{\Text(75,40)[l]{$v_2$}}
    \LightBlue{\Text(40,75)[b]{$e_1$}}
    \Cyan{\Text(40,45)[b]{$e_2^{}$}}
    
  \end{axopicture}
}
\hspace{1cm}
\stackrel{*}{\longrightarrow}
\hspace{1cm}
\raisebox{-15pt}{
  \begin{axopicture}(80,50)(0,35)
    \SetWidth{1.0}
    \SetColor{Black}
	\LightBlue{\DashArc[arrow,clockwise](40,40)(30,180,0){3}}
	\Cyan{\Line[arrow](70,40)(10,40)}
    \Green{\Vertex(10,40){2}}
    \Red{\Vertex(70,40){2}}
    \Green{\Text(5,40)[r]{$v_1'$}}
    \Red{\Text(75,40)[l]{$v_2'$}}
    \LightBlue{\Text(40,75)[b]{$e_1'$}}
    \Cyan{\Text(40,45)[b]{$e_2'$}}
  \end{axopicture}
  
}
\label{eq:yukcc}
\ee 
Again, we code the labels by colors. The two vertices of the graph are complex conjugates 
of each other and hence receive different labels (red and green). Notice that these labels    change under complex conjugation due to the reversal of the arrows. The label-preserving map $v_1\to v_2'$, $v_2\to v_1'$ is an isomorphism  and shows that the graph is self-conjugate.
Taking the complex conjugate of the explicit expression for $\Gamma$, using eqns.~(\ref{eq:scalarherm}) and (\ref{eq:fermpropreal})
\begin{multline}
\left(\tr\left[
B( {\t_1}; {x_1},{x_2})
\bar C({x_2})
(i{\sl D_{ 2}}+i{\sl \k_2}+{m_2})B( {\t_2}; {x_2},{x_1})
C({x_1})
\right]e^{-{m_1^2\t_1}+{m_2^2\t_2}} \right)^*
\\
=\tr\left[
B( {\t_1}; {x_2},{x_1})
\bar C({x_1})
(i{\sl D_{ 1}}+i{\sl \k_2}+{m_2})B( {\t_2}; {x_1},{x_2})
C({x_2})
\right]e^{-{m_1^2\t_1}+{m_2^2\t_2}}\,,
\end{multline}
shows that the second line is indeed the expression we would obtain from the graph on the right hand side of eq.~(\ref{eq:yukcc}), confirming eq.~(\ref{eq:ccGamma}).

\section{Conclusions}
\label{sec:conclusions}

We have elaborated on the recently proposed covariant perturbation theory formalism of ref.~\cite{vonGersdorff:2022kwj}, casting the contribution of each multi-loop Feynman graph into a  factorized form, eq.~(\ref{eq:master}), which allowed us to perform the momentum integral in closed form. The latter is expressed in terms of one polynomial $\Delta$ and three matrices $\Q$, $\R$, and $\U$.  The latter, when multiplied by $\Delta$, are also polynomial in the Schwinger-Feynman parameters. We have expressed them both in terms of the graph's incidence matrix and as sums over subgraphs. The two representations are complementary in that they highlight different properties of the four quantities. We remark that the parametrization in terms of the incidence matrix is more suitable for a possible future automation of the formalism.

Our findings can be applied to standard perturbation theory for momentum space correlation functions as explained in sec.~\ref{sec:Dirac}, bypassing the usual 
 tensor reduction algorithms.

We furthermore have discussed automorphisms of graphs and shown that they leave each factor of eq.~(\ref{eq:master}) separately invariant. 

\acknowledgments
I would like to thank Kevin Santos and Sylvain Fichet for discussions.
I acknowledge financial support by the Conselho Nacional de Desenvolvimento Científico e Tecnológico (CNPq) under fellowship number 309448/2020-4. 

\appendix

\section{Cycle spaces}

\label{app:cycle}

In this appendix we briefly discuss the kernel of the incidence matrix $\B$ and the closely related concept of cycle space.

Consider a graph $G$ and a spanning tree $S$. Adding back in any of the $L$ deleted edges gives a one-loop connected graph, let
 us denote this graph with $C$ and the corresponding incidence matrix with $\B_C$, which is (being connected and one loop)  a $V\times V$ matrix of rank $V-1$.
Therefore, its adjugate (a.k.a.~cofactor transposed) matrix $(\B_C)^{\rm adj}$ has rank 1, see eq.~(\ref{eq:ranks}).
Since we know that the cokernel of $\B_C$ is spanned by $(1,1,\dots 1)$, and from $\B_C(\B_C)^{\rm adj}=0=(\B_C)^{\rm adj}\B_C$, 
we must have that 
\be
\B_C^{\rm adj}=
v_C^{}
(1,1,\dots 1)\,,
\ee
where the column vector $v_{C}$ spans the one dimensional kernel of $\B_{C}$, describing  the single loop of $C$. It remains to fix the normalization of $v_C$.
By a known result in graph theory (see for instance Lemma 2.6 in ref.~\cite{bapat2010graphs}), the determinants of the incidence matrix of a connected tree graph with one row deleted are all $\pm 1$.
Therefore the entries of $(\B_C)^{\rm adj}$ and hence the $(v_C)_i$ are $\pm 1$ when $i$ participates in the loop and zero otherwise. 
It is clear from the form of $\B_C$  that when the edges $i$ and $j$ are oriented in the same way in the loop they have the same sign, and if they are oppositely oriented they have the opposite sign
(otherwise $v_C$ would not be a vector of the kernel of $\B_C$). 

This provides a formal way of finding a basis for the null space of $\B$ (by choosing an arbitrary spanning tree, adding in deleted edges one at a time, and computing the adjugate matrices). There are $L$ different possibilities to add in back deleted edges, the $v_C$ are called the fundamental cycles of $S$. 
Clearly, they are linearly independent (they have one edge unique to each of them) and hence the $v_C$ form a (non-orthonormal) basis for $\ker \B$. They can be normalized  so all entries equal $\pm 1$.\footnote{One can of course also read off these fundamental cycles  by inspection of the graph $C$. Adding in an edge to $S$ to form $C$ connects two vertices, these two vertices are already connected by a unique path in $S$. The cycle is given by joining the path and the edge, and $v_C$ can be read off immediately.
}

The $v_C$ obtained this way are called the simple cycles of $G$ (simple referring to the fact that the corresponding $C$ are one-loop.  Here is a list of all simple cycles of the 
graph of figure \ref{fig:threeloop}.
\be
\begin{pmatrix}
1\\1\\0\\0\\0\\0
\end{pmatrix},
\begin{pmatrix}
0\\1\\1\\1\\1\\0
\end{pmatrix},
\begin{pmatrix}
0\\0\\0\\0\\1\\1
\end{pmatrix},
\begin{pmatrix}
1\\0\\-1\\-1\\-1\\0
\end{pmatrix},
\begin{pmatrix}
0\\1\\1\\1\\0\\-1
\end{pmatrix},
\begin{pmatrix}
1\\0\\-1\\-1\\0\\1
\end{pmatrix}.
\label{eq:simplecycles}
\ee
Any three of them are linearly independent and form a basis for $\ker \B$.

A note on terminology. In graph theory the term "cycle space"  usually refers to  the kernel of $\B $ over the two-element field $\mathbb Z_2$. This space has dimension $L$ and consists of $2^L$ vectors (for instance, the complete cycle space of the graph in figure \ref{fig:threeloop} consists of the vectors in eq.~(\ref{eq:simplecycles}) modulo 2, as well as the zero-vector and the vector $(1,1,0,0,1,1)\tra$).
Throughout this work,  cycle space  means $\ker \B$ (over the reals), and we will always  use the term cycle to refer to elements of the cycle space normalized such that its entries equal $\pm 1$.

The alliances defined in section \ref{sec:char1} can be identified by looking at any cycle basis.   
Two edges $e_i,\ e_j$ that are not bridges are allies iff for every element $\c$ in the basis $\c_i=\c_j \mod 2$.
Removing $e_i$ and $e_j$ from the graph reduces the dimension of the cycle space by one (it must decrease because $e_i$ and $e_j$ are not bridges, and 
it cannot decrease by more than one because in this case a cycle would exist that contained one edge but not the other). Using $P'=P-2$, $L'=L-1$, $V'=V$, we have 
$C'=C+1$, hence the number of disconnected components increased by one and $\{e_i,e_j\}$ form a bond. Conversely, if $\{e_i,e_j\}$ form a bond, removing these edges results in $C'=C+1$ and hence $L'=L-1$ which implies that there cannot be any cycle that contains one edge but not the other, as otherwise we would have had $L'=L-2$. This shows equivalence of the two characterizations of alliances in section \ref{sec:char1}.
\footnote{Let $b$ be the vector that has a one at positions $i$ and $j$ and zeroes elsewhere.
Then the condition can be rephrased as $b\tra c=0\mod 2$ for all basis cycles $c$. In other words, $b$ is an element of the orthogonal compliment of the cycle space over $\mathbb Z_2$, which is also called bond space \cite{Bondy1976}. It is isomorphic to the row space of the incidence matrix over $\mathbb Z_2$.}

Let us now prove that eq.~(\ref{eq:Qbb}) follows from eq.~(\ref{eq:Qvv}). We need to show that for any given simple cycle $\c$, 
\be
\sum_{\substack{C\\\overline v_C=\c}}\omega_C=\Delta_\c
\ee
Fix any $i$ with $c_i\neq 0$, that is, any edge that participates in $\c$. Denote by $j\neq i$ the remaining edges of $\c$. Then the graphs $C$ are in one to one correspondence with the spanning trees $S$ that do not contain $i$ but contain all $j$,
(the correspondence being removing/adding edge $i$). Hence
\be
\sum_{\substack{C\\\overline v_C=\c}}\omega_C
=\sum_{\substack{S\\ i\notin S,j\in S}}\frac{\omega_S}{\t_i}
=\sum_S [\partial_{\t_i}\omega_S]_{\t_j=0}
=[ \partial_{\t_i}\D]_{\t_j=0}
\ee
Owing to the contraction and deletion relations \cite{Weinzierl:2022eaz}, this amounts to  first deleting edge $i$ and then contracting edges $j$. But this is the same as contracting the whole loop.

\section{Moore-Penrose pseudo inverses}
\label{sec:MP}
Consider an $N\times M$ real-valued matrix $\M$ with possibly nontrivial kernel and cokernel. 
The Moore-Penrose pseudo inverse of $M$ is defined via the conditions \cite{Penrose:1955vy}
\be
\M\M^+\M=\M\,,\qquad
\M^+\M\M^+=\M^+\,,\qquad
(\M^+\M)\tra=\M^+\M\,,\qquad
(\M\M^+)\tra=\M\M^+\,.
\label{eq:MP}
\ee
These conditions fix $\M^+$ uniquely. 
An equivalent formulation can be given in terms of projectors onto column and row spaces of $\M$, denoted here by $\Pr_{\im \M}$ and $\Pr_{\im \M\tra}$ respectively.
Either of the following pairs of equations is equivalent to the four equations in eq.~(\ref{eq:MP})
\bea
\M^+\M=\Pr_{\im \M\tra}\,,\qquad 
\M^+=\M^+\Pr_{\im \M}\,.
\label{eq:MP2}
\eea
\bea
\M\M^+=\Pr_{\im \M}\,,\qquad 
\M^+=\Pr_{\im \M\tra}\M^+\,.
\label{eq:MP3}
\eea
{\em Proof.}
We will first show that each of (\ref{eq:MP}), (\ref{eq:MP2}), (\ref{eq:MP3}) defines a unique matrix.
Uniqueness of $\M^+$ from eq.~(\ref{eq:MP}) is a classic result \cite{Penrose:1955vy}. To see that eq.~(\ref{eq:MP2}) fixes $\M^+$ uniquely, consider any $v\in \im \M$. By standard linear algebra, there exists a unique $w\in \im \M\tra$ with $v=\M w$. For two matrices $\ms X,\ms Y$ satisfying the relations in eq.~(\ref{eq:MP2}), calculate $\ms Xv=\ms X\M w=w$, $\ms Yv=\ms Y\M w=w$, and hence $\ms Xv=\ms Yv$ for all $v\in \im \ms M$, and hence $\ms X\Pr_{\im \M}=\ms Y\Pr_{\im \M}$. The second relation in eq.~(\ref{eq:MP2}) then implies $\ms X=\ms Y$.
The uniqueness from eq.~(\ref{eq:MP3}) is analogous. 
To see existence of $\M^+$ and to prove that all three definitions give the same $\M^+$ one  considers the singular value decomposition of $\M$ \cite{strang2016introduction}
\be
\M=\ms O_L \begin{pmatrix}
\M_{\rm diag} & {0}\\
0&0
\end{pmatrix}
\ms O_R\tra\,,
\ee
where $\ms M_{\rm diag}$ is the diagonal matrix containing the $\operatorname{rank} \M$ nonzero singular values of $\M$,  the zeroes stand for rectangular matrices of the appropriate dimensions, and $\ms O_{L,R}$ are orthogonal matrices.
One then defines the $M\times N$ matrix \cite{strang2016introduction}
\be
\M^+\equiv 
\ms O_R \begin{pmatrix}
\ms M^{-1}_{\rm diag} &0\\
0&0
\end{pmatrix}
\ms O_L\tra\,.
\ee
It is a trivial matter to show that $\M^+$ defined in this way satisfies all equations in eq.~(\ref{eq:MP}), (\ref{eq:MP2}) and (\ref{eq:MP3}), and hence they all define the same matrix $\M^+$ and are therefore equivalent characterizations of the MP inverse, {\em q.e.d.}

The characterizations eq.~(\ref{eq:MP2}) and eq.~(\ref{eq:MP3}) are useful in practice. For instance consider {\em any} left pseudo-inverse $\ms X$ of $\M$, i.e., any matrix that satisfies the first relation in eq.~(\ref{eq:MP2}). Then $\M^+=\ms X \Pr_{\im \M}$. 

Consider now the case of trivial kernel, but nontrivial cokernel (an injective but not surjective map).
It is easily seen that $\M\tra \M$ is regular, and that $(\M\tra \M)^{-1}\M\tra$ is a left inverse for $\M$. 
It is readily checked by either eq.~(\ref{eq:MP}) or eq.~(\ref{eq:MP2}) that it is equal to the MP inverse $\M^+$. 

In eq.~(\ref{eq:kS}) we wrote  particular left inverses for the incidence matrix $\B_S$ of some connected tree graph $S$. 
We now simply multiply any of these left inverses with $\Pr_{\im \B_S}$ from the right to get the Moore-Penrose pseudo inverse. Let us chose $k^1_{S,i}=p_{T_i^1}$, and call this left inverse $\ms X$. Let us assume that without loss of generality the ingoing momenta are the first $V_i^1$ vertices. Then the $i$th row of the left inverse $\ms X$ reads $(-1,-1\dots ,-1,0,
\dots,0)$. The projector reads 
\be
\Pr_{\im \B_S}=
\one -
\tfrac{1}{V}
(1,\dots, 1)\tra(1,\dots, 1)\,,
\ee
and hence
\be
(-1,-1\dots ,-1,0,\dots,0)\Pr_{\im \B_S}=
(\tfrac{-V^2_i}{V},\dots \tfrac{-V^2_i}{V},\tfrac{V^1_i}{V},\dots \tfrac{V^1_i}{V})\,,
\ee
is the $i$th row of the matrix $\B_S^+$.

\bibliographystyle{JHEP}

\bibliography{literature}

\providecommand{\href}[2]{#2}\begingroup\raggedright\begin{thebibliography}{10}

\bibitem{schwinger1951}
J.~Schwinger, {\it On gauge invariance and vacuum polarization},  {\em Phys.
  Rev.} {\bf 82} (Jun, 1951) 664--679.

\bibitem{dewitt1965dynamical}
B.~DeWitt, {\em Dynamical Theory of Groups and Fields}.
\newblock Documents on modern physics. Gordon and Breach, 1965.

\bibitem{DeWitt:1967ub}
B.~S. DeWitt, {\it {Quantum Theory of Gravity. 2. The Manifestly Covariant
  Theory}},  {\em Phys. Rev.} {\bf 162} (1967) 1195--1239.

\bibitem{Gilkey:1975iq}
P.~B. Gilkey, {\it {The Spectral geometry of a Riemannian manifold}},  {\em J.
  Diff. Geom.} {\bf 10} (1975), no.~4 601--618.

\bibitem{Barvinsky:1985an}
A.~O. Barvinsky and G.~A. Vilkovisky, {\it {The Generalized Schwinger-Dewitt
  Technique in Gauge Theories and Quantum Gravity}},  {\em Phys. Rept.} {\bf
  119} (1985) 1--74.

\bibitem{Avramidi:1990ug}
I.~G. Avramidi, {\it {The Covariant technique for the calculation of the heat
  kernel asymptotic expansion}},  {\em Phys. Lett. B} {\bf 238} (1990) 92--97.

\bibitem{Avramidi:1990je}
I.~G. Avramidi, {\it {The Covariant Technique for Calculation of One Loop
  Effective Action}},  {\em Nucl. Phys. B} {\bf 355} (1991) 712--754. [Erratum:
  Nucl.Phys.B 509, 557--558 (1998)].

\bibitem{Fujikawa:1979ay}
K.~Fujikawa, {\it {Path Integral Measure for Gauge Invariant Fermion
  Theories}},  {\em Phys. Rev. Lett.} {\bf 42} (1979) 1195--1198.

\bibitem{Fujikawa:1980eg}
K.~Fujikawa, {\it {Path Integral for Gauge Theories with Fermions}},  {\em
  Phys. Rev. D} {\bf 21} (1980) 2848. [Erratum: Phys.Rev.D 22, 1499 (1980)].

\bibitem{Ball:1988xg}
R.~D. Ball, {\it {Chiral Gauge Theory}},  {\em Phys. Rept.} {\bf 182} (1989) 1.

\bibitem{Schmidt:1993rk}
M.~G. Schmidt and C.~Schubert, {\it {On the calculation of effective actions by
  string methods}},  {\em Phys. Lett. B} {\bf 318} (1993) 438--446,
  [\href{http://arxiv.org/abs/hep-th/9309055}{{\tt hep-th/9309055}}].

\bibitem{vonGersdorff:2003dt}
G.~von Gersdorff and M.~Quiros, {\it {Localized anomalies in orbifold gauge
  theories}},  {\em Phys. Rev. D} {\bf 68} (2003) 105002,
  [\href{http://arxiv.org/abs/hep-th/0305024}{{\tt hep-th/0305024}}].

\bibitem{vonGersdorff:2006nt}
G.~von Gersdorff, {\it {Anomalies on Six Dimensional Orbifolds}},  {\em JHEP}
  {\bf 03} (2007) 083, [\href{http://arxiv.org/abs/hep-th/0612212}{{\tt
  hep-th/0612212}}].

\bibitem{Hoover:2005uf}
D.~Hoover and C.~P. Burgess, {\it {Ultraviolet sensitivity in higher
  dimensions}},  {\em JHEP} {\bf 01} (2006) 058,
  [\href{http://arxiv.org/abs/hep-th/0507293}{{\tt hep-th/0507293}}].

\bibitem{Barvinsky:2005qi}
A.~O. Barvinsky, {\it {The Gospel according to DeWitt revisited: Quantum
  effective action in braneworld models}},  in {\em {International Conference
  on Theoretical Physics Dedicated to the 70 Year Anniversary of the Tamm
  Theory Department}}, 4, 2005.
\newblock \href{http://arxiv.org/abs/hep-th/0504205}{{\tt hep-th/0504205}}.

\bibitem{vonGersdorff:2008df}
G.~von Gersdorff, {\it {One-Loop Effective Action in Orbifold
  Compactifications}},  {\em JHEP} {\bf 08} (2008) 097,
  [\href{http://arxiv.org/abs/0805.4542}{{\tt arXiv:0805.4542}}].

\bibitem{Henning:2014wua}
B.~Henning, X.~Lu, and H.~Murayama, {\it {How to use the Standard Model
  effective field theory}},  {\em JHEP} {\bf 01} (2016) 023,
  [\href{http://arxiv.org/abs/1412.1837}{{\tt arXiv:1412.1837}}].

\bibitem{Drozd:2015rsp}
A.~Drozd, J.~Ellis, J.~Quevillon, and T.~You, {\it {The Universal One-Loop
  Effective Action}},  {\em JHEP} {\bf 03} (2016) 180,
  [\href{http://arxiv.org/abs/1512.03003}{{\tt arXiv:1512.03003}}].

\bibitem{delAguila:2016zcb}
F.~del Aguila, Z.~Kunszt, and J.~Santiago, {\it {One-loop effective lagrangians
  after matching}},  {\em Eur. Phys. J. C} {\bf 76} (2016), no.~5 244,
  [\href{http://arxiv.org/abs/1602.00126}{{\tt arXiv:1602.00126}}].

\bibitem{Henning:2016lyp}
B.~Henning, X.~Lu, and H.~Murayama, {\it {One-loop Matching and Running with
  Covariant Derivative Expansion}},  {\em JHEP} {\bf 01} (2018) 123,
  [\href{http://arxiv.org/abs/1604.01019}{{\tt arXiv:1604.01019}}].

\bibitem{Zhang:2016pja}
Z.~Zhang, {\it {Covariant diagrams for one-loop matching}},  {\em JHEP} {\bf
  05} (2017) 152, [\href{http://arxiv.org/abs/1610.00710}{{\tt
  arXiv:1610.00710}}].

\bibitem{FuentesMartin:2016uol}
J.~Fuentes-Martin, J.~Portoles, and P.~Ruiz-Femenia, {\it {Integrating out
  heavy particles with functional methods: a simplified framework}},  {\em
  JHEP} {\bf 09} (2016) 156, [\href{http://arxiv.org/abs/1607.02142}{{\tt
  arXiv:1607.02142}}].

\bibitem{Ellis:2017jns}
S.~A.~R. Ellis, J.~Quevillon, T.~You, and Z.~Zhang, {\it {Extending the
  Universal One-Loop Effective Action: Heavy-Light Coefficients}},  {\em JHEP}
  {\bf 08} (2017) 054, [\href{http://arxiv.org/abs/1706.07765}{{\tt
  arXiv:1706.07765}}].

\bibitem{Kramer:2019fwz}
M.~Kr\"amer, B.~Summ, and A.~Voigt, {\it {Completing the scalar and fermionic
  Universal One-Loop Effective Action}},  {\em JHEP} {\bf 01} (2020) 079,
  [\href{http://arxiv.org/abs/1908.04798}{{\tt arXiv:1908.04798}}].

\bibitem{Ellis:2020ivx}
S.~A.~R. Ellis, J.~Quevillon, P.~N.~H. Vuong, T.~You, and Z.~Zhang, {\it {The
  Fermionic Universal One-Loop Effective Action}},  {\em JHEP} {\bf 11} (2020)
  078, [\href{http://arxiv.org/abs/2006.16260}{{\tt arXiv:2006.16260}}].

\bibitem{Angelescu:2020yzf}
A.~Angelescu and P.~Huang, {\it {Integrating Out New Fermions at One Loop}},
  {\em JHEP} {\bf 01} (2021) 049, [\href{http://arxiv.org/abs/2006.16532}{{\tt
  arXiv:2006.16532}}].

\bibitem{Cohen:2020fcu}
T.~Cohen, X.~Lu, and Z.~Zhang, {\it {Functional Prescription for EFT
  Matching}},  {\em JHEP} {\bf 02} (2021) 228,
  [\href{http://arxiv.org/abs/2011.02484}{{\tt arXiv:2011.02484}}].

\bibitem{Dittmaier:2021fls}
S.~Dittmaier, S.~Schuhmacher, and M.~Stahlhofen, {\it {Integrating out heavy
  fields in the path integral using the background-field method: general
  formalism}},  {\em Eur. Phys. J. C} {\bf 81} (2021), no.~9 826,
  [\href{http://arxiv.org/abs/2102.12020}{{\tt arXiv:2102.12020}}].

\bibitem{Cohen:2023gap}
T.~Cohen, X.~Lu, and Z.~Zhang, {\it {Anomaly Cancellation in Effective Field
  Theories From the Covariant Derivative Expansion}},
  \href{http://arxiv.org/abs/2301.00827}{{\tt arXiv:2301.00827}}.

\bibitem{Cohen:2023hmq}
T.~Cohen, X.~Lu, and Z.~Zhang, {\it {Anomalies from the covariant derivative
  expansion}},  {\em Phys. Rev. D} {\bf 107} (2023), no.~11 116015,
  [\href{http://arxiv.org/abs/2301.00821}{{\tt arXiv:2301.00821}}].

\bibitem{Larue:2023uyv}
R.~Larue and J.~Quevillon, {\it {The Universal One-Loop Effective Action with
  Gravity}},  \href{http://arxiv.org/abs/2303.10203}{{\tt arXiv:2303.10203}}.

\bibitem{Duff:1975ue}
M.~J. Duff and M.~Ramon-Medrano, {\it {On the Effective Lagrangian for the
  Yang-Mills Field}},  {\em Phys. Rev. D} {\bf 12} (1975) 3357.

\bibitem{Batalin:1976uv}
I.~A. Batalin, S.~G. Matinyan, and G.~K. Savvidy, {\it {Vacuum Polarization by
  a Source-Free Gauge Field}},  {\em Sov. J. Nucl. Phys.} {\bf 26} (1977) 214.

\bibitem{Batalin:1978gt}
I.~A. Batalin and G.~K. Savvidy, {\it {Vacuum Polarization by Covariant
  Constant Gauge Field Two Loop Approximation}}, .

\bibitem{Bornsen:2002hh}
J.~P. Bornsen and A.~E.~M. van~de Ven, {\it {Three loop Yang-Mills beta
  function via the covariant background field method}},  {\em Nucl. Phys. B}
  {\bf 657} (2003) 257--303, [\href{http://arxiv.org/abs/hep-th/0211246}{{\tt
  hep-th/0211246}}].

\bibitem{Vassilevich:2003xt}
D.~V. Vassilevich, {\it {Heat kernel expansion: User's manual}},  {\em Phys.
  Rept.} {\bf 388} (2003) 279--360,
  [\href{http://arxiv.org/abs/hep-th/0306138}{{\tt hep-th/0306138}}].

\bibitem{vonGersdorff:2022kwj}
G.~von Gersdorff and K.~Santos, {\it {New covariant Feynman rules for effective
  field theories}},  {\em JHEP} {\bf 04} (2023) 025,
  [\href{http://arxiv.org/abs/2212.07451}{{\tt arXiv:2212.07451}}].

\bibitem{Fliegner:1997rk}
D.~Fliegner, P.~Haberl, M.~G. Schmidt, and C.~Schubert, {\it {The Higher
  derivative expansion of the effective action by the string inspired method.
  Part 2}},  {\em Annals Phys.} {\bf 264} (1998) 51--74,
  [\href{http://arxiv.org/abs/hep-th/9707189}{{\tt hep-th/9707189}}].

\bibitem{Weinzierl:2022eaz}
S.~Weinzierl, {\em {Feynman Integrals}}.
\newblock 1, 2022.

\bibitem{Bogner:2010kv}
C.~Bogner and S.~Weinzierl, {\it {Feynman graph polynomials}},  {\em Int. J.
  Mod. Phys. A} {\bf 25} (2010) 2585--2618,
  [\href{http://arxiv.org/abs/1002.3458}{{\tt arXiv:1002.3458}}].

\bibitem{Golz_2017}
M.~Golz, {\it New graph polynomials in parametric {QED} feynman integrals},
  {\em Annals of Physics} {\bf 385} (oct, 2017) 328--346.

\bibitem{bapat2010graphs}
R.~Bapat, {\em Graphs and Matrices}.
\newblock Universitext. Springer London, 2010.

\bibitem{Bondy1976}
J.~A. Bondy and U.~S.~R. Murty, {\em Graph Theory with Applications}.
\newblock Elsevier, New York, 1976.

\bibitem{Penrose:1955vy}
R.~Penrose, {\it {A Generalized inverse for matrices}},  {\em Proc. Cambridge
  Phil. Soc.} {\bf 51} (1955) 406--413.

\bibitem{strang2016introduction}
G.~Strang, {\em Introduction to Linear Algebra}.
\newblock Wellesley, 2016.

\end{thebibliography}\endgroup

\end{document}